\documentclass[aps,twocolumn,showpacs,notitlepage,reprint,preprintnumbers,superscriptaddress,amsmath,amssymb,nofootinbib]{revtex4-1}
\usepackage{graphicx}
\usepackage{dcolumn}
\usepackage{bm}
\usepackage{setspace}
\usepackage{epstopdf}
\usepackage{amsmath}
\usepackage{enumerate}
\usepackage{tabularx}
\usepackage[usenames,dvipsnames,svgnames,table]{xcolor}
\usepackage[colorlinks=true,linkcolor=gray,urlcolor=blue,citecolor=blue]{hyperref}
\usepackage{ifthen}		

\def\simge{\mathrel{
     \rlap{\raise 0.511ex \hbox{$>$}}{\lower 0.511ex \hbox{$\sim$}}}}
\def\simle{\mathrel{
     \rlap{\raise 0.511ex \hbox{$<$}}{\lower 0.511ex \hbox{$\sim$}}}}
\def\be{\begin{equation}}
\def\ee{\end{equation}}
\def\bea{\begin{eqnarray}}
\def\eea{\end{eqnarray}}

\newcommand{\calo}{{\cal O}}

\newcommand{\Tr}{{\rm Tr}\,}
\newcommand{\ReTr}{{\rm ReTr}\,}
\renewcommand{\Re}{{\rm Re}\,}

\newcommand{\eq}[1]{\begin{eqnarray}#1\end{eqnarray}}  
\newcommand{\eqa}[1]{\begin{align}#1\end{align}}
\newcommand{\eqal}[2]{\begin{align}\label{#2}#1\end{align}}
 
\newcommand{\eql}[2]{\begin{eqnarray}\label{#2}#1\end{eqnarray}}  

\newcommand{\ax}{a_x}
\newcommand{\axd}{a_x^\dagger}
\newcommand{\bx}{b_x}
\newcommand{\bxd}{b_x^\dagger}
\newcommand{\ay}{a_y}

\newcommand{\by}{b_y}

\newcommand{\JLU}{Institut f\"ur Theoretische Physik,
  Justus-Liebig-Universit\"at, 35392 Giessen, Germany} 

\newcommand{\RU}{Institut f\"ur Theoretische Physik, Universit\"at Regensburg, 93053 Regensburg, Germany}

\begin{document}
\singlespacing
\title{A Hybrid-Monte-Carlo study of monolayer graphene with partially
  screened\\  Coulomb interactions at finite spin density}
\author{Michael K\"orner}
\affiliation{\JLU}

\author{Dominik Smith}
\affiliation{\JLU}

\author{Pavel Buividovich}
\affiliation{\RU}

\author{Maksim Ulybyshev}
\affiliation{\RU}

\author{Lorenz von Smekal}
\affiliation{\JLU}

\date{\today}
\begin{abstract}
\noindent 
We report on Hybrid-Monte-Carlo simulations at finite spin density of
the $\pi$-band electrons in monolayer graphene with realistic
inter-electron interactions. Unlike simulations at finite charge-carrier   
density, these are not affected by a fermion-sign problem.  
Our results are in qualitative agreement with an interaction-induced
warping of the Fermi contours, and a reduction of the bandwidth as
observed in angle resolved photoemission spectroscopy 
experiments on charge-doped graphene systems. Furthermore, we find
evidence that the neck-disrupting Lifshitz transition, which occurs
when the Fermi level traverses the van Hove singularity (VHS), becomes a
true quantum phase transition due to interactions. This is in-line 
with an instability of the VHS towards the formation of electronic
ordered phases, which has been predicted by a variety of different
theoretical approaches. 
\end{abstract}

\maketitle

\section{Introduction}
\begin{figure}
\begin{center}
\includegraphics[width=0.97\linewidth]{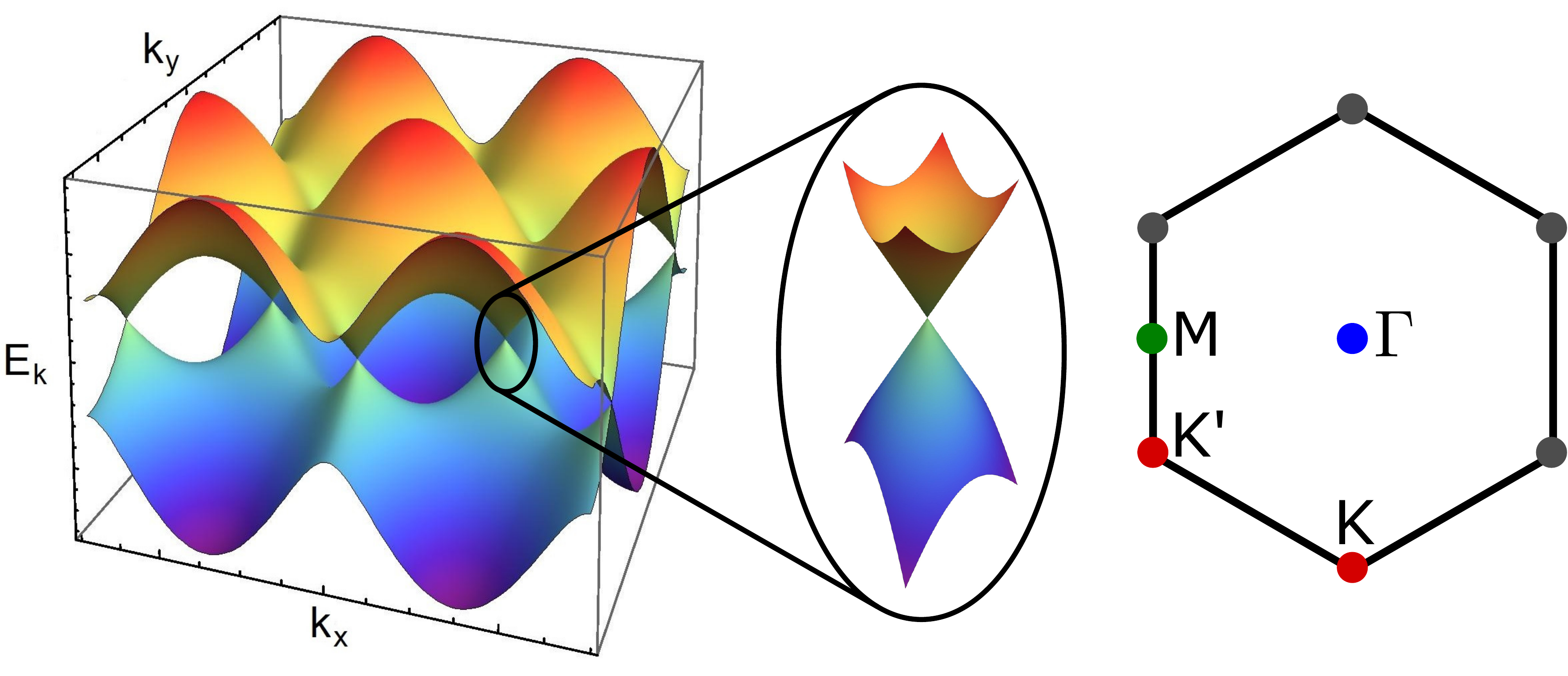}
\caption{Left: Electronic band structure of the nearest-neighbor tight-binding
theory of graphene. Dirac cones around the K-points are enlarged. Right: The first Brillouin zone
and terminology for special points therein.\label{fig:tbbands}}
\end{center}
\end{figure}

Already the nearest-neighbor hexagonal tight-binding model \cite{Wallace:1947an}
qualitatively captures many of the interesting features of
monolayer graphene, such as the existence of 
massless electronic excitations near the corners of the first
Brillouin zone (K-points) with a linear dispersion relation for the
low-energy excitations around those Dirac points \cite{Gusynin:2007ix}.  
In the electronic bands one also finds saddle points, located
at the M-points, which are characterized by a vanishing group velocity. 
These separate the low energy region, described by an effective Dirac
theory, from a region where electronic quasi-particles behave like a regular Fermi liquid with a parabolic
dispersion relation centered around the  $\Gamma$-points. See Fig.~\ref{fig:tbbands} for an illustration of the valence and conduction bands of the nearest-neighbor tight-binding theory. 

\begin{figure}
\begin{center}
\includegraphics[width=0.97\linewidth]{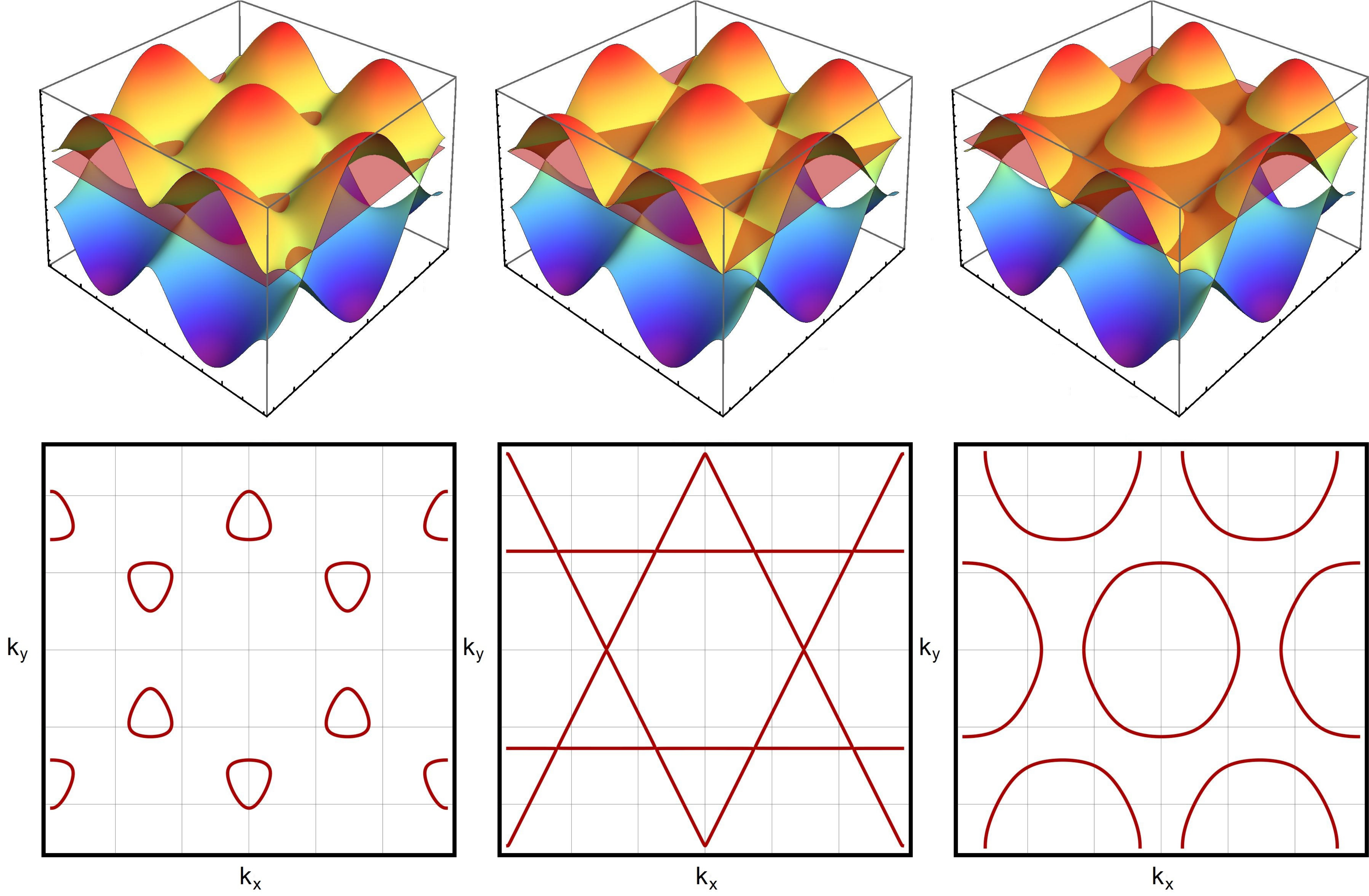}
\caption{ Topology of the Fermi lines (intersection lines with
horizontal planes) for Fermi levels below (left), exactly at (middle) and above (right) the saddle points.
\label{fig:Lifshitz1}}
\end{center}
\end{figure}

When the Fermi level is shifted across the saddle points by a chemical potential, a change of the
topology of the Fermi surface (which is one-dimensional for a 2D crystal)
takes place. The distinct circular Fermi (isofrequency) lines surrounding the Dirac points are deformed into triangles
when the saddle point is approached, meet to form one large connected region and then break up again
into circles around the $\Gamma$-points (see
Fig.~\ref{fig:Lifshitz1}). This is known as neck-disrupting  
Lifshitz transition \cite{Lifshitz:1960su}.

The Lifshitz transition is not a true phase transition in the
thermodynamic sense (as it is purely topological and not associated with any type of spontaneous symmetry breaking i.e. formation of
an ordered phase), but exhibits features commonly associated with
such: singularities in free energy and susceptibility at zero
temperature with the chemical potential as the control
parameter. Unlike phase transitions these singularities are
logarithmic (in two dimensions) and not due to interactions but 
to the vanishing group velocity of electronic excitations at
the saddle points which leads to a logarithmic divergence 
in the density of states (DOS) with increasing surface area of the graphene
sheet. This is known as a van Hove singularity (VHS)
\cite{VanHove:1953kkj} and can be 
observed in a pure form, for instance, in microwave photonic crystals
with a Dirac spectrum as macroscopic models for the non-interacting
graphene band structure \cite{Dietz:2013sga,Dietz:2016aj} and
fullerenes with an Atiyah-Singer index theorem
\cite{Dietz:2015cna}. 
  
The fate of the VHS of monolayer graphene in the presence of many-body interactions is a topic of active research.
Since interactions are strongly enhanced by the divergent DOS, it is generally believed that
the VHS is unstable towards formation of electronic ordered phases. This would imply that the 
Lifshitz transition becomes a true phase transition in a realistic description of the interacting
system at sufficiently low temperatures. It is known that superconductivity can arise from purely repulsive interactions through
the Kohn-Luttinger mechanism \cite{Kohn:1965ss}. Furthermore, it is known that VHSs exist close
to the Fermi level in most high-$T_c$ superconducting cuprates, so it has long been discussed whether 
they produce superconducting instabilities generically (known as the ``van Hove scenario'' \cite{Markiewicz:1997zh}).
This scenario was also proposed for doped graphene \cite{Bostwick:2010as}. 
An exciting possibility specific to graphene furthermore is the emergence of an anomalous time-reversal symmetry violating 
chiral d-wave superconducting phase from electron-electron repulsion close to the VHS 
\cite{Gonzalez:2008hn,Chubukov:2012as,PhysRevB.86.020507,Kagan:2014gf,Schaffer:2014sd,Loethman:2014gg, Wang:2012as}.

The theoretical perspective is not unambiguous, however. The underlying reason is that several
competing channels exist for interaction-driven instabilities at the VHS, and that a subtle
interplay of different mechanisms (nesting of the Fermi surface and deviations thereof,
relative interaction strengths of couplings at different distances, accounting for
electron-phonon interactions etc.) can tilt the balance towards
one phase or another. Aside from d-wave superconductivity different formalisms have, for example, 
predicted superconductivity with pairing in a channel of f-wave symmetry \cite{Gonzalez:2013op}, 
spin-density wave (SDW) phases \cite{Makogon:2011da}, a Pomeranchuk instability 
\cite{Valenzuela:2008fa,Lamas:2009cv} or a Kekul\'{e} superconducting pattern \cite{Faye:2016xx}.  And this is by no means an exhaustive list.

On the experimental side, by now there exist several techniques to shift the Fermi level of graphene to the van Hove singularity: The VHS can be probed in systems where gold nanoclusters  
are intercalated between monolayer graphene and epitaxal graphene \cite{Cranney:2010ad}, by chemical 
doping \cite{McChesney:2007uh,Bostwick:2010as}, by gating \cite{Novoselov:2005df,Zhang:2005ff,Efetov:2010sb}
or in ``twisted graphene'' \cite{Li:2010sd} (stacked graphene layers with a rotation angle). 
Furthermore the valence and conduction bands of graphene can be precisely mapped
using angle resolved photoemission spectroscopy (ARPES). Such experiments show
clear evidence for a reshaping of the graphene bands by many-body interactions \cite{Bostwick200763}
and for a warping of the Fermi surface, leading to an extended, not pointlike, 
van Hove singularity (EVHS) characterized by the flatness of the bands, i.e. lack of
energy dispersion, along one direction \cite{Bostwick:2010as}.\footnote{This is a rather general phenomenon which can also exist, e.g. around the saddle points in the dispersion relation of a triangular lattice \cite{ext_vanHove2}. It is considered to be a crucial mechanism in the context of the ``van Hove scenario'', since it enhances the singularity in the DOS and thus possible instabilities towards ordered phases, such as superconductivity.} ARPES experiments on many different doped graphene systems have also shown bandwidth renormalizations 
with deviations of several $100$ meV from single-particle band models \cite{Ulstrup:2016ha}
and a massive enhancement of the electron-phonon coupling at the VHS \cite{McChesney:2007uh}.
Unambiguously distinguishing different electronic phases close to the VHS however is an open
experimental challenge. 

In this work, results of Hybrid-Monte-Carlo (HMC) simulations of the
interacting tight-binding theory of graphene are presented. These
simulations were carried out at finite chemical potential for spin
rather than charge density, as induced by a spin-staggered chemical
potential. Although the effects of the two are substantially
different, both kinds of chemical  potential can be used 
to tune Fermi levels across the entire range of the $\pi$-bands,
including the VHS. The only difference, however substantial, is that
the spin-staggered chemical potential shifts the Fermi levels of the two
spin orientations in opposite directions corresponding to the pure
Zeeman splitting of an in-plane magnetic field \cite{Aleiner:2007va}.

Technically this modification is necessary to avoid the
fermion-sign problem which otherwise arises from the complex phase of
the fermion determinant in the charge-doped system, and which causes
importance sampling to break down. The system with spin-staggered
chemical potential may be viewed as the so-called ``phase-quenched''
version (defined by the modulus of the fermion determinant in the
 measure) of graphene at finite charge density. Because the two
spin components of the $\pi$-band electrons in graphene correspond to
two different fermion flavors, this is entirely analogous to
simulating two-flavor QCD at finite isospin density with pion
condensation rather than finite baryon density in the form of self-bound
nuclear matter which is equally impossible due to a strong sign problem. The
phases are clearly distinct but many important questions and genuine
finite-density effects in lattice simulations can be addressed at
finite isospin density as well.

The particular questions addressed here are about the genuine effects of
inter-electron interactions on the VHS and the Lifshitz transition in
graphene. Our main focus thereby is the behavior of susceptibilities
to identify signatures of instabilities and phase transitions. To
directly study the interaction-driven instabilities that might occur
in the charge-doped systems described above would require us to
measure the particle-hole susceptibility at finite charge density
which is however not possible due to the sign
problem. We therefore simulate at finite spin density and measure the
susceptibility corresponding to ferromagnetic spin-density
fluctuations instead which does not have this problem.
In the non-interacting limit the two agree, and either
one may be used to characterize the electronic Lifshitz transition.
Because the spin-staggered chemical potential used here
could at least in principle be realized in experiment as well, by
sufficiently strong in-plane magnetic fields, our study might also
become relevant in its own right in the future. 

We chose a realistic microscopic inter-electron interaction
potential which accounts for screening by electrons in the
$\sigma$-bands \cite{Wehling:2011df}. A 
range of different system sizes and temperatures were considered
(these are temperatures of the electron gas only, as our simulations
presently do not account for phonons). Furthermore, the inter-electron
interaction potential was rescaled to different magnitudes, ranging
from zero to the full interaction strength of suspended graphene.
 
The purpose of this work is two-fold: First
we wish to assess whether the effects of interactions on the VHS at
finite spin density can at least qualitatively be compared with the
observations from ARPES data at finite charge density. To this end, we
study the reshaping of the $\pi$-bands of the interacting system (with
respect to a ``flattening'' scenario). Secondly, we want to exemplify 
how the logarithmic divergence of a susceptibility at the VHS in the
$T \to 0$ limit can change to a critical scaling law at non-zero $T_c$
in the presence of inter-electron interactions, as this would signal
the existence of an electronic ordered state close to the VHS and indicate  
that the Lifshitz transition becomes a true quantum phase transition
(with $\mu$ as a control parameter) below this $T_c$. Identifying the
precise nature of the ordered phase will of course depend on the choice of
chemical potential and is thus beyond the scope of this work,
however. 

This paper is structured as follows: In the following chapter we
discuss the behavior of the particle-hole susceptibility in the
non-interacting tight-binding theory with temperature and system
size where it agrees with that of the ferromagnetic spin-density
fluctuations. Exact results for the non-interacting system
will serve as a baseline for our studies of the effects of
inter-electron interactions. 
As the HMC method necessitates the introduction of a non-zero
temperature of the electron gas (due to the introduction of a
Euclidean time dimension which must be of finite extent) and  of
finite system size, the derivation accounts for both. Furthermore, we
derive the leading temperature dependence at the VHS, of the divergent
peak height of the susceptibility,  
in the infinite volume limit. In Chapter \ref{sec:Setup} the
Hybrid-Monte-Carlo simulation of the interacting theory is introduced,
with emphasis on the fermion-sign problem which arises at finite 
 chemical potential for charge-carrier density. We derive expressions
 for the ferromagnetic and antiferromagnetic spin-density 
 susceptibilities expressed in terms of the inverse fermion matrix. In
Chapter \ref{sec:Results} results of the HMC calculations are
presented. These include detailed studies of the temperature and
interaction-dependent behavior of the ferromagnetic susceptibility
with particular emphasis on the fate of the VHS.  Preliminary results
concerning the possibility of spin-density wave order from the
corresponding antiferromagnetic susceptibility are also presented. We
then provide our summary and conclusions in Chapter \ref{sec:Conclusion}.

\section{Particle-hole susceptibility and Lifshitz transition}

\subsection{Non-interacting tight-binding theory}
\label{sec:TBLI}

As mentioned in the introduction, in the nearest-neighbor
tight-binding description of the $\pi$-bands in graphene, due to
particle-hole symmetry the particle-hole susceptibility is independent
of the sign of the chemical potential $\mu $. Because this is true
independently for both spin components, there is thus no distinction
between the susceptibilities for charge and spin fluctuations in the
non-interacting case, and both equally reflect the Lifshitz transition
at finite charge or spin density. The chemical potential $\mu $  this
section can therefore be used for either one interchangeably. 

In order to understand the relation between the VHS in the electronic
quasi-particle  DOS $\rho(\omega)$, the Thomas-Fermi susceptibility
$\chi $  and the properties of the neck-disrupting electronic Lifshitz
transition, one best starts from the particle-hole polarization
function $\Pi(\omega, \vec{p}; \mu, T)$ at temperature $T$ and chemical
potential $\mu $ for charge-carrier density (with $\mu =0 $ at half
filling), excitation frequency $\omega $ and momentum $\vec p$.   

The particle-hole polarization function determines the charge-density correlations corresponding to the diagonal time component of the polarization tensor in QED. Using the imaginary-time formalism and subsequent analytic continuation with the appropriate boundary conditions for retarded Green's functions, at one-loop one arrives at the expression,
\eqal{\Pi(\omega, \vec{p}; \mu, T) =& -  \int_\mathrm{BZ} \frac{d^2k}{(2\pi)^2} \sum_{s,s' = \pm 1 }  \\ 
  & \hskip -1cm  \frac{g_\sigma}{2} \bigg( 1   + ss' \frac{\Re \big( \phi^*_{\vec{k}}\,\phi_{\vec{k}+\vec{p}}\big)}{|\phi_{\vec{k}}| |\phi_{\vec{k}+\vec{p}}|} \bigg)  \notag \\
  & \times
\frac{n_f\big( \beta(s'\epsilon_{\vec{k}+\vec{p}} - \mu ) \big) - n_f\big( \beta (s \epsilon_{\vec{k}} - \mu) \big)}{ s'\epsilon_{\vec{k}+\vec{p}}- s\epsilon_{\vec{k}}-\omega -i\epsilon}\, , \notag }{chiint}
where $g_\sigma = 2 $ here for the spin degeneracy,
$\phi_{\vec{k}}=\sum_n e^{i\vec{k}\vec{\delta}_n}$ is the structure factor with 
nearest-neighbor vectors  $\vec\delta_n$, $n= 1,2,3$ on the hexagonal lattice,  and single-particle energies  $\epsilon_{\vec k} = \kappa|\phi_{\vec{k}}|$ (where $\kappa $ is the hopping parameter) in Fermi-Dirac distributions $n_f(x) = 1/(e^{x}+1) $ at $\beta = 1/T$. 

The particle-hole polarization or Lindhard function $\Pi$ is a sum of terms describing particle-hole excitations within the same band for $s'=s$ (intraband), and terms describing interband excitations for $s' = -s$. The complete one-loop expressions for intraband and interband transitions have been computed from Eq.~(\ref{chiint}) in closed analytic form in Refs.~\cite{Stauber:2010mn,Dietz:2013sga}. 

\begin{figure}
\begin{center}
\includegraphics[width=0.97\linewidth]{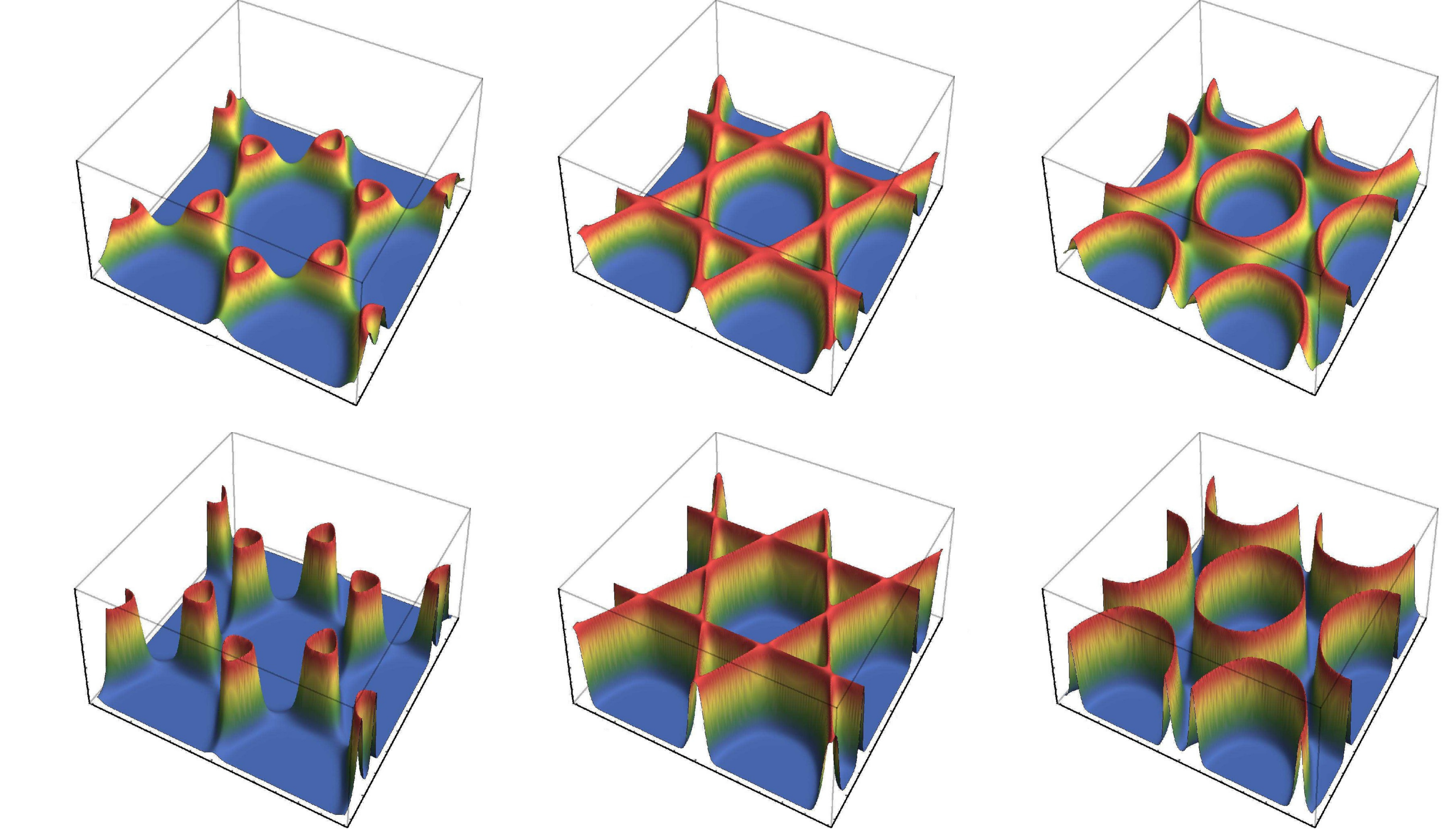}
\caption{ Integrand of Eq.~(\ref{eq:CHITEMP}) for values of $\mu$
  below (right), at (middle) and above (left) the van Hove singularity; 
from the top to the bottom row the temperature has been lowered
by a factor $1/2$ (from $T = \kappa/2$ to $\kappa/4$). 
\label{FINITE_T}}
\end{center}
\end{figure}

The imaginary parts of $\Pi$ vanish in the limit $\omega\to 0$ which describes  
static Lindhard screening. In a subsequent long-wavelength limit $\vec p\to 0$, to which only interband excitations contribute, one obtains the usual Thomas-Fermi susceptibility, 
\eql{\chi(\mu) =  A_c \, \lim_{\vec{p} \rightarrow 0} \, \lim_{\omega \rightarrow 0} \Pi(\omega, \vec{p}; \mu, T)\, ,}{eq:TFs} 
here normalized per unit cell of area $A_c = 3 \sqrt 3 a^2/2$ with nearest-neighbor distance $a \approx 1.42 \text{\AA}$ for the carbon atoms in graphene. It is straightforwardly calculated as
\eqal{\chi(\mu)  =&  \frac{g_\sigma A_c}{4T} \int_\mathrm{BZ} \frac{d^2k}{(2\pi)^2}  \\
  & 
\times\left[ \text{sech}^{2}\left( \frac{\epsilon_{\vec{k}}-\mu }{2T} \right) 
+ \text{sech}^{2}\left( \frac{\epsilon_{\vec{k}}+\mu }{2T} \right) \right]~.\notag}{eq:CHITEMP}
With the present normalization, the zero-temperature limit of $\chi(\mu)$ then in turn agrees with the density of states per unit cell $\rho(\epsilon) $ at the Fermi level $\epsilon =\mu$, i.e.
\eql{ \lim_{T\to 0} \, \chi(\mu) = g_\sigma A_c \int_\mathrm{BZ}
  \frac{d^2k}{(2\pi)^2} \,  \delta(\epsilon_{\vec{k}}-|\mu|) \equiv
  \rho(\mu)~. }{eq:dosdef}
Fig.~\ref{FINITE_T} demonstrates explicitly how the integrand in
(\ref{eq:CHITEMP}) encodes the effect of temperature on the
susceptibility. The sharp Fermi lines which were shown in the lower
row of Fig.~\ref{fig:Lifshitz1} are smeared out, since a spread of
different energy levels may now be excited. The allowed range becomes
narrower as temperature is lowered and concentrates on the 
Fermi level with $\chi $ approaching the DOS there, for $T\to 0 $,
cf.~Eq.~(\ref{eq:dosdef}).  

The density of states was first derived for transverse 
vibrations of a hexagonal lattice by Hobson and Nierenberg
in 1953 \cite{Hobson:1953df}. They found logarithmic divergences near
the saddles of the energy bands, i.e., the van Hove singularities, as
well as the zeros now identified with the Dirac points. From the
corresponding analytical expression of the hexagonal tight-binding model given in \cite{CastroNeto:2009zz}, one readily obtains for the fermionic system at
finite charge-carrier density, with a Fermi energy near one of the van Hove
singularities at $\mu  = \pm \kappa$, 
\begin{equation}
\rho(\mu)  = \frac{3g_\sigma }{2\pi^2\kappa} \Big\{  - \frac{1}{2}
\ln\Big(\frac{|\mu |}{\kappa} - 1\Big)^2 + 2\ln 2 + \mathcal
O\Big(\frac{|\mu|}{\kappa} -1\Big) \Big\}  \, . \label{vHs}
\end{equation}
The correspondingly diverging zero-temperature
susceptibility $\chi$ is due to the infinite degeneracy of 
ground states of the two-dimensional fermionic system when the Fermi
level passes through the van Hove singularity. In the thermodynamic
sense this can be considered as a zero temperature transition with
control parameter $|\mu|$.  To illustrate this one    
introduces the reduced Fermi-energy parameter $z 
= (|\mu|-\kappa)/\kappa$ to rewrite (\ref{vHs}),
\begin{equation}
\chi(z)  = \frac{3 g_\sigma }{2\pi^2\kappa} \Big(  - 
\ln |z|  + 2\ln 2 + \mathcal O(z) \Big)  \, . \label{LT}
\end{equation}
Unlike the cases of first or second order phase transitions, the
susceptibility does not diverge with a power law but
logarithmically. This is a manifestation of the neck-disrupting electronic
Lifshitz transition in two dimensions
\cite{Lifshitz:1960su,Blanter1994}. 
There is no obvious change in symmetry, the
transition is only due to the topology change of the Fermi surface. The
singular part of the corresponding thermodynamic grand potential
is non-zero on both sides of the transition. The original argument is
simple, one expands the 
single-particle energy around a saddle point at $\kappa$
in suitable coordinates,
\begin{equation}
  \epsilon_{\vec k} = \kappa + \frac{k_x^2}{2m_1} - \frac{k_y^2}{2m_2}
  \, , \label{eq:saddle_ni}
\end{equation}
which gives in Eq.~(\ref{eq:dosdef}) a singular contribution
\begin{equation}
  \rho_s(z) = - \frac{g_\sigma A_c}{2\pi^2} \, \sqrt{m_1 m_2} \,
  \ln|z| \, . 
\end{equation}
For the nearest-neighbor tight-binding model on the hexagonal
lattice, one verifies that  $\sqrt{m_1 m_2} = 1/(\kappa A_c )  $ so that
$\rho_s(z) = - g_\sigma/(2\pi^2 \kappa) \ln |z|$. With a factor of 3 for the
three $M$ points per Brillouin zone this agrees with the leading
behavior of the zero-temperature susceptibility in Eq.~(\ref{LT}) as
it should.
One integration over $\kappa z$ then yields the number of states
in an interval around the saddle, a second one the corresponding
contribution to the grand potential $\Omega$ per unit cell
which hence acquires a corresponding singularity \cite{Blanter1994}   
\begin{equation}
  \Omega_\mathrm{sing} = \frac{3 g_\sigma \kappa }{2\pi^2} \,
  \frac{z^2}{2} \ln|z| \, .  
\end{equation}
It is symmetric around $z=0$. There is thus no order parameter in the
usual sense, but one may discuss this transition in terms of a change
in the approximate symmetries of 
the low-energy excitation spectrum with some analogy in excited-state quantum
phase transitions \cite{Dietz:2013sga}.

At any rate, the logarithmic singularity of the electronic Lifshitz
transition in the grand potential is restricted to strictly zero
temperature. To see this explicitly, we first use the density of
states to express the
finite-temperature susceptibility in the following form,
\eqal{\chi(\mu)  =&  \frac{1}{4T} \int_{0}^{3\kappa} d\epsilon \, \rho(\epsilon) \,\notag \\& \times
\left[ \text{sech}^{2}\left( \frac{\epsilon-\mu}{2T} \right) + \text{sech}^{2}\left( \frac{\epsilon+\mu}{2T} \right) \right]~.}{eq:chiintegral2} 
Assuming $\mu > 0$ for now, we may drop the second term in the
brackets for sufficiently low temperatures, and extend the limits of
integration to $\pm \infty$. For the susceptibility maximum at $\mu =
\kappa $ we can furthermore approximate $\rho(\epsilon)$ by the
expansion in Eq.~(\ref{vHs}) in the region of support of the integrand
around $\epsilon = \kappa $ to obtain,
\eql{\hspace*{-.4cm}\chi_\mathrm{max}  = \frac{3g_\sigma }{2\pi^2\kappa} 
\Big\{ -  \ln{\big({\pi T}/{\kappa} \big)} + \gamma_E  + 3 \ln
2 + \mathcal{O}(T) \Big\}}{Temppeakgl}
where $\gamma_E$ is the Euler-Mascheroni constant. The maximum of the
susceptibility of the electronic Lifshitz transition is finite at
finite $T$.

In this way, the logarithmic divergence in the DOS at the VHS is reflected
in the Thomas-Fermi susceptibility $\chi(\mu)$. At low but finite
temperatures $\chi(\mu)$ peaks when the Fermi level crosses the VHS
(for $\mu = \kappa $ in the non-interacting system). The peak
height grows logarithmically as temperature is lowered. Its
divergence in the zero-temperature limit is a manifestation of the
neck-disrupting electronic Lifshitz transition with its logarithmic
singularity in the chemical potential as the corresponding control
parameter. 

So much for the non-interacting and infinite system. Before we discuss
finite volume effects and interactions, we can speculate how a
reshaping of the saddle points in the single-particle band structure
by interactions might qualitatively affect the Lifshitz
transition. If we assume a non-Fermi liquid behavior near the saddles
for example of the form
\begin{equation}
  \epsilon_{\vec k} = \epsilon_0 + \kappa \big( c_1 (k_x a)^\alpha -
  c_2 (k_ya)^\alpha\big)\, , \label{saddlereshape}
\end{equation}
instead of (\ref{eq:saddle_ni}), where we had $\sqrt{c_1 c_2} = 3\sqrt 3/4$,
$\epsilon_0 = \kappa $ and $\alpha =2$ for the non-interacting
tight-binding model, we now obtain analogously,  
\begin{equation}
  \rho_s(z) \propto \kappa^{-1} |z|^{-\gamma} \, , \;\;\mbox{with}\;\;
  \gamma = 1- \frac{2}{\alpha} \, .
\end{equation}
In Eq.~(\ref{eq:chiintegral2}) this for $\mu = \epsilon_0$ then readily yields
\begin{equation}
  \chi_\mathrm{max} \propto \frac{1}{\kappa} \,
  \Big(\frac{\kappa}{T}\Big)^\gamma \, ,
\end{equation}
replacing Eq.~(\ref{Temppeakgl}) for $\gamma\not=0$. We can see that,
e.g.\ for $\alpha = 4$ in single-particle energies near the saddles
(\ref{saddlereshape}), the logarithmic divergence of Eq.~(\ref{Temppeakgl})
turns into a square root divergence of the susceptibility  maximum for
$T\to 0$  with $\gamma = 1/2$. Whereas the limit of a completely flat
single-particle energy band with $\alpha \to\infty $ would correspond to
$\gamma = 1 $ and hence $\chi_\mathrm{max} \propto 1/T$.

We conclude this section by reiterating that for vanishing two-body interactions,  $\chi(\mu)$ is blind to a change of sign. And this is
true for each of the spin orientations separately. We will use
opposite signs of $\mu$ for the two spin orientations in our
simulations below to avoid a fermion-sign problem. While this then
corresponds to a Zeeman splitting, as caused by an in-plane magnetic
field for example, rather than a change of the charge-carrier density
away from half filling, the tight-binding results are unaffected by
such a sign change. We may therefore thus use $\chi(\mu)$ with
unlike-sign chemical potentials for the two spin states, analogous to
isospin chemical potential in Quantum Chromodynamics (QCD), to detect
deviations from the pure tight-binding theory in our
Hybrid-Monte-Carlo (HMC) simulations, where it can be readily obtained
(discussed in Sec.~\ref{subsec:Observables}).  

To make the comparison between the Lifshitz transition in the
non-interacting system and the results from HMC simulations with interactions as
direct as possible, in the next subsection  we first derive  
semi-analytic expression for $\chi(\mu)$ in the tight-binding model on
finite lattices with the same boundary conditions that we use in the
simulations.

\subsection{Finite lattices}
In our HMC simulations we study graphene sheets of finite surface area, with 
periodic boundary conditions along the primitive vectors 
$\vec{a}_{1,2}=\frac{a}{2} (\sqrt{3}, \pm 3)$ 
(where $a \approx 1.42 \text{\AA}$ is the inter-atomic distance on the hexagonal lattice)
spanning one of the triangular sub-lattices
(``Born-von Karman boundary conditions''). 
We simulate symmetric lattices, with $N$ unit cells
along each axis. To take finite size into account, Eq.~(\ref{eq:CHITEMP}) 
is rewritten as a sum over the allowed momentum states, 
which are given by the Laue condition $e^{i\vec{k}\vec{R}} = 1$, where $\vec{R}=n\vec{a}_1+m\vec{a}_2$ with $n,m \in [1,\cdots,N]$. The momentum states are
\eql{\vec{k}= \frac{n}{N}\vec{b}_1 + \frac{m}{N}\vec{b}_2~,}{eq:lauevectors}
where $\vec{b}_{1,2} = \frac{2\pi}{3a} (\sqrt{3}, \pm 1) $ are the the
base vectors of the reciprocal lattice. The integral measure $d^{2}k$
turns into a finite surface element $(\Delta k)^2= | \vec{b}_1
\times \vec{b}_2 |/N^2 =A_\mathrm{BZ}/N^2$, where $A_\mathrm{BZ} =
(2\pi)^2/A_c$ is the area of the first Brillouin zone,  
and the integral in Eq.~(\ref{eq:CHITEMP}) for the susceptibility of a
finite sheet becomes, 
\eqal{\chi(\mu) =  \frac{g_\sigma}{ 4 T N^2 }   \sum_{n,m}&   \left[
    \text{sech}^{2}\left( \frac{\epsilon_{mn} - \mu }{2T} \right)
    \right. \notag \\ &
    \hskip 1cm \left. + \, \text{sech}^{2}\left(
    \frac{\epsilon_{mn} +\mu}{2T} \right) \right]~. }{eq:chisum}   
Here $\epsilon_{mn} $ is the dispersion relation, evaluated at the points defined by Eq.~(\ref{eq:lauevectors}):
\eqa{\epsilon_{mn} 
  = \kappa \left\{3 + 4\cos{ \left(\pi\frac{n+m}{N}\right)}\cos{\left(\pi \frac{n-m}{N}\right)}\right.\notag\\ \left. + 2\cos{\left(2\pi\frac{n+m}{N}\right)} \right\}^{\frac{1}{2}}~. }
Eq.~(\ref{eq:chisum}) is of a form which can be compared directly to
the simulations. The sums cannot be carried out analytically, but are
straightforward to evaluate numerically. 

\begin{figure}
\begin{center}
\includegraphics[width=0.97\linewidth]{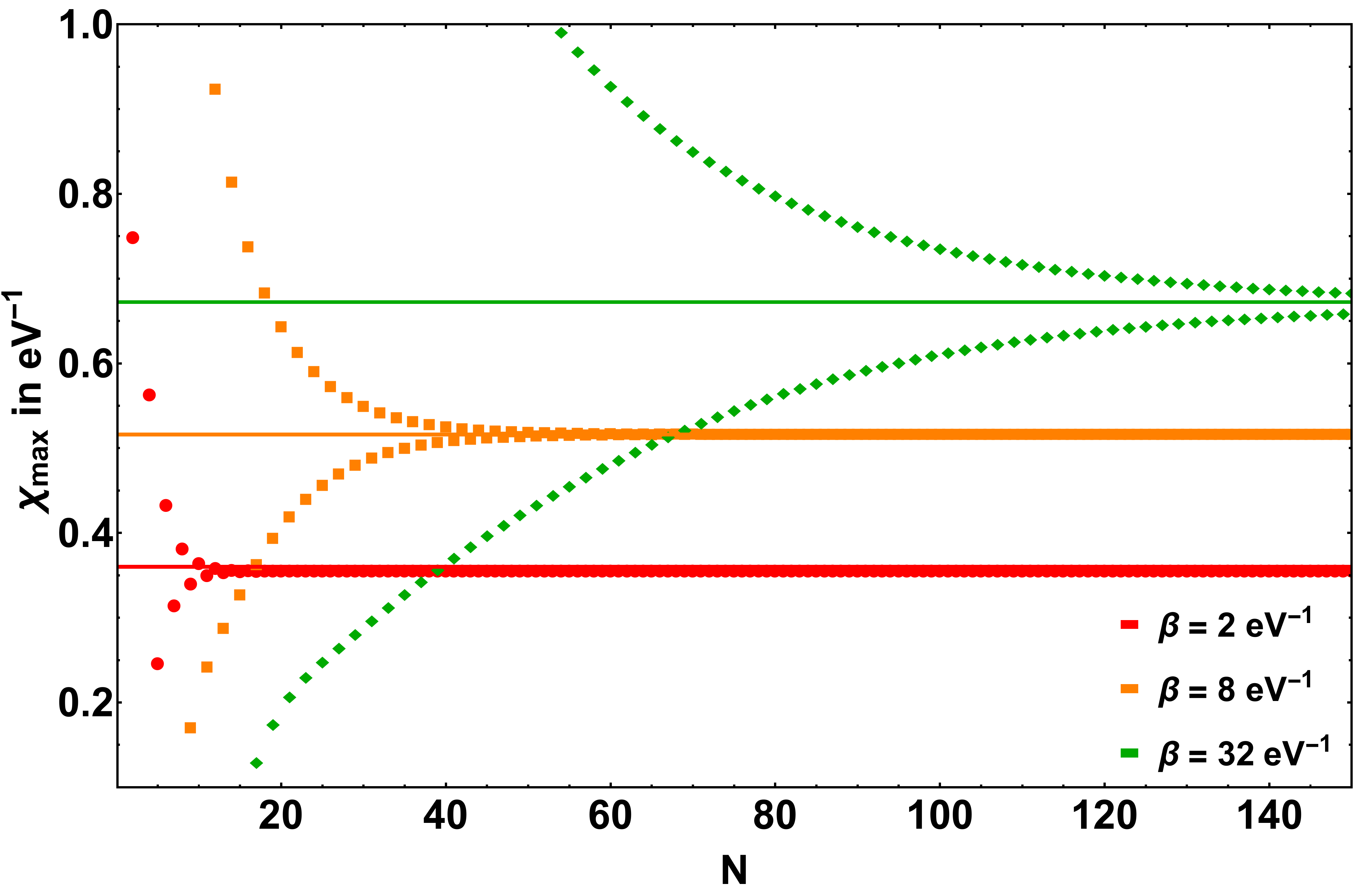}
\caption{Finite-size scaling of the susceptibility peak 
  at different temperatures ($\beta=1/T$) from
  Eq.~(\protect\ref{eq:chisum}); the horizontal lines indicate the
  leading-order prediction from Eq.~(\protect\ref{Temppeakgl}), the slight
  deviations of the infinite volume limit from this prediction for
  $\beta = 2$ eV$^{-1}$ are due to $\mathcal O(T)$ corrections.
\label{DIFF_TEMP_OVER_VOL}}
\end{center}
\end{figure}

Of course there is no  divergence of the particle-hole susceptibility 
in a finite volume, not even at zero temperature. The spectrum is
discrete and the total number
of states is finite, so the density of states cannot diverge either. 
In Ref.~\cite{Dietz:2013sga} it was shown, however, that the finite-size scaling
of the susceptibility maximum at $T=0$ is logarithmic likewise, namely
\begin{equation}
  \chi_\mathrm{max} = \frac{3g_\sigma}{2\pi^2\kappa } \, \Big(
  \ln N_c - 2 \ln \pi +1 +\calo(1/N_c) \Big) \, , \label{eq:lifshitzscaling}
\end{equation}
where $N_c = N^2$ is the number of unit cells. Since our simulations
are carried out at finite temperature, it is clear that we cannot
observe this behavior directly because it is valid only at strictly
zero temperature. The extension of the analytic expressions to finite
volume and finite temperature is not so straightforward, however, and cannot be
done analytically. 

Therefore, we use the implicit representation of $\chi(\mu) $ for a
finite sheet at temperature $T$ in Eq.~(\ref{eq:chisum}) and compute
the sums numerically. The results of $\chi$  at $\mu=\kappa$  are
shown for various lattice sizes and temperatures in
Fig.~\ref{DIFF_TEMP_OVER_VOL}. In general, for any 
finite temperature, $\chi(\mu=\kappa,N)$ for $N \to \infty$ approaches
a flat asymptote $\propto  \ln(\beta\kappa ) $ which in turn increases
with $\beta = 1/T$ according to Eq.~(\ref{Temppeakgl}). 
It is the temperature dependence of these asymptotic values 
which follows Eq.~(\ref{Temppeakgl}). Convergence to the infinite
volume limit becomes slower for decreasing temperatures as the
asymptotic value increases. 

Fig.~\ref{DIFF_TEMP_OVER_VOL} shows a strong influence of the parity of the lattice, where odd
lattices approach the $N \to \infty$ limit from below and even lattices from above. For a 
fixed lattice size, the peak height either diverges (for even lattices)
or goes to zero (for odd lattices) as $T\to 0$. This difference
arises from the fact that the sums in Eq.~(\ref{eq:chisum}) only
contain momentum modes which hit the M-points exactly
when $N$ is even. For even $N$, points on the lines with
$\text{sech}^2 ((\epsilon_{mn}-\mu)/2T) =1 $ contribute with diverging
weight $\propto 1/T$  to the sum 
(cf.~Fig.~\ref{FINITE_T}), while for odd $N$ there are no such points
but only points that cluster around these lines when the system
becomes large. 

\section{Inter-electron interactions}

\subsection{Simulation setup}
\label{sec:Setup}
The present work implements Hybrid-Monte-Carlo simulations of the
interacting tight-binding theory on the 
hexagonal graphene lattice, based on a formalism developed by Brower
et al.~\cite{Brower:2012zd, Brower:2012ze}, 
which goes beyond the low-energy approximation 
(studied extensively in the past 
\cite{Drut:2008rg,Drut:2009gg,Drut:2009rf,Drut:2010ff,Drut:2011fd,Hands:2008dg,Armour:2009vj,Armour:2011ff,Buividovich:2012uk,Giedt:2011df})
and is thus able to capture the full band structure beyond the Dirac cones. The HMC method on the graphene lattice is by now well established, and has been successfully applied in conclusive studies of the antiferromagnetic phase 
transition \cite{Buividovich:2012nx, Ulybyshev:2013swa, Smith:2013pxa, Smith:2014tha, Smith:2014vta}
as well as in ongoing studies of the phase diagram of an extended fermionic Hubbard model on the hexagonal graphene lattice \cite{Buividovich:2016tgo}.
Numerous other topics were also addressed with HMC, such as the optical conductivity of graphene \cite{Boyda:2016emg},
the effect of hydrogen adatoms
\cite{Ulybyshev:2015opa,Buividovich:2017sju} or the single
quasi-particle spectrum of carbon nanotubes \cite{Luu:2015gpl}. 

We have written about our setup in great detail in the past (see ref. \cite{Smith:2014tha} for a step-by-step derivation) and
will only provide a summary here. In particular, we focus on the additional challenges which arise when introducing
a chemical potential (i.e.~the fermion-sign problem) and discuss our workaround solution (a spin-dependent sign flip). 
To be clear: This work does not attempt to solve the sign-problem, but rather studies a modified Hamiltonian which is free
of such a problem. To assess to what degree the physics is changed by this
modification is part of the motivation for this work. 

The starting point is the interacting tight-binding Hamiltonian in second-quantized form
\begin{align}
H =& \sum_{\langle x,y \rangle}(-\kappa)(\axd \ay + \bxd \by + \textrm{h.c.}) \notag\\
&+ \sum_{x,y}\, q_x V_{xy} q_y +  \sum_{x}m_s (\axd \ax + \bxd \bx)~.
\label{eq:tightbinding}
\end{align}
The chemical potential is absent at this stage and will be introduced later.
The first sum in Eq.~(\ref{eq:tightbinding}) runs over pairs of
nearest neighbors only (with a hopping parameter $\kappa=2.7 $ eV), so
we neglect higher order hoppings.
The other sums run over all sites (including both sublattices) of the $2D$ hexagonal lattice. 
Here $\axd,\ax$ denote creation/annihilation operators for electrons
in the $\pi$-bands with spin $+1/2$ in the $z$-direction
(perpendicular to the graphene sheet) and $\bxd,\bx$ are analogous
operators for holes (``anti-particles'')
with spin $-1/2$. The hopping term also contains a sublattice
dependent sign-flip for the $\bxd,\bx$ operators
\cite{Smith:2014tha}. 

We have also added in Eq.~(\ref{eq:tightbinding}) a staggered mass term
$m_s= (-1)^s \, m$ with a sublattice $s=0,1$ dependent sign to regulate the
low-lying eigenvalues of the Hamiltonian, as is customary in
lattice-QCD simulations. While simulations at exactly zero mass are possible
in principle  \cite{Buividovich:2016tgo} (unlike lattice QCD there
appear to be no topological obstructions to simulating at exactly zero
mass here), a finite mass term has numerical advantages, and   
it only affects the low-lying excitations around the 
Dirac points which are not the primary focus of our present study. In
fact, our investigation of the Lifshitz transition turns out to be
rather insensitive to this mass term as one might expect, based on the
band structure of the non-interacting system, as long as $m_s \ll \kappa $. 
Moreover, a spin and sublattice-staggered mass term of this form
also serves as an external field for sublattice-symmetry breaking by
spin-density wave formation. So derivatives 
with respect to $m_s$ may be used to detect an instability of the ground 
state towards SDW order. 

The operator $q_x=\axd\ax - \bxd \bx$ represents physical charge. Interactions are taken to be
instantaneous, which is true to good approximation since $v_F \ll c$, where $v_F$ is the Fermi velocity
of the electrons. One of the great advantages 
of the instantaneous Hamiltonian in Eq.~(\ref{eq:tightbinding})
(compared to implementing the photon as an Abelian gauge field on  
link variables) is that any positive-definite matrix can be chosen for $V_{xy}$,
leaving a great freedom to choose a realistic two-body potential to describe microscopic interactions.
In particular, it is possible to implement deviations from pure
Coulomb-type interactions due to screening from $\sigma$-band and
other localized electrons. 

\begin{figure}
\vspace{4mm}
\begin{center}
\includegraphics[width=0.97\linewidth]{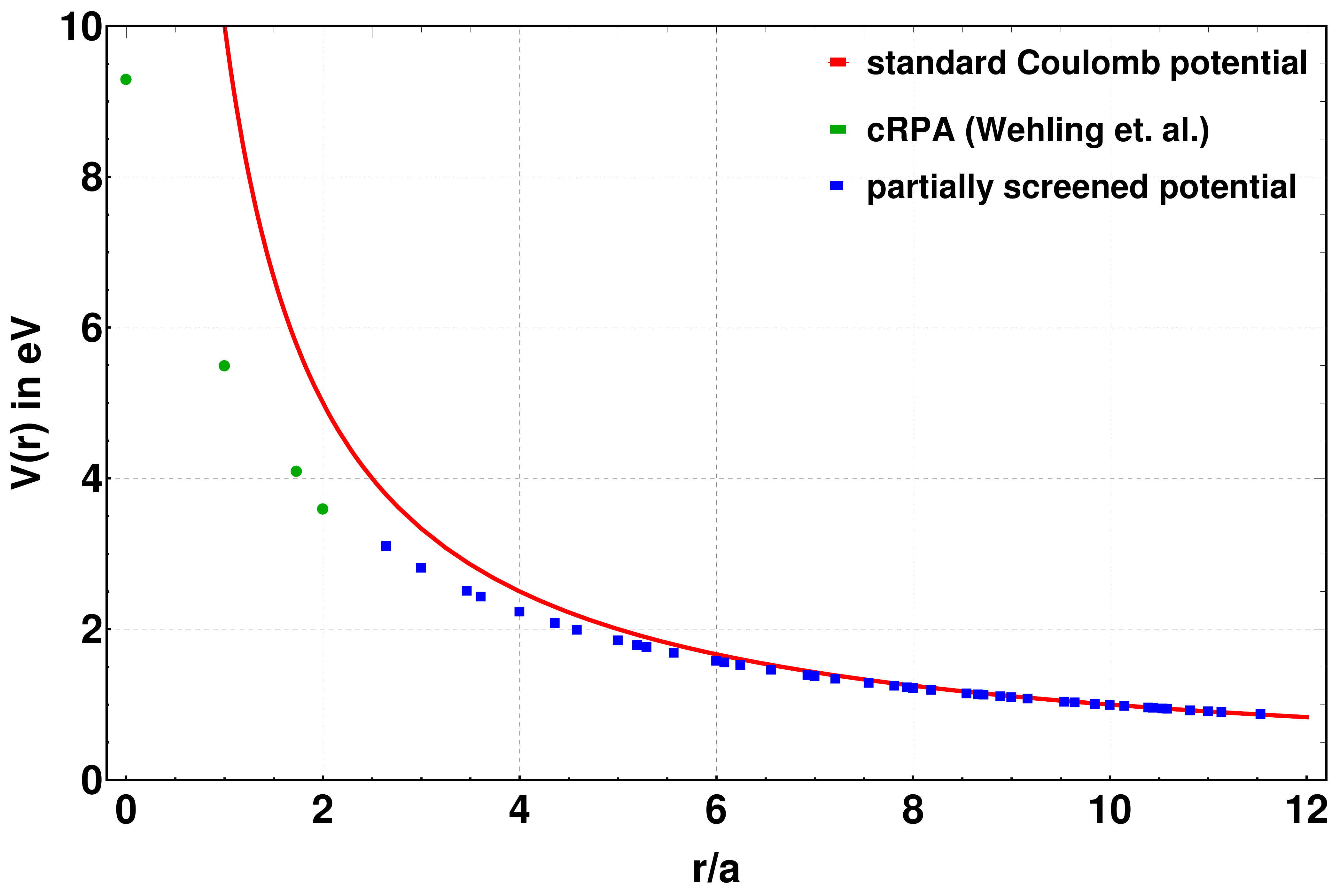}
\caption{Comparison of the standard Coulomb
    potential (red) with the partially screened potential given by
    Eq.~(\ref{eq:potfit}). The first four points are exact cRPA
    results of Ref.~\cite{Wehling:2011df} (green), while the remaining
    ones are obtained from the interpolation based on the thin-film
    model from the same reference (blue).}  
\label{fig:potential}
\end{center}
\end{figure}

In this work, we choose a two-body potential which accounts for precisely
this screening as obtained from calculations within a constrained random-phase
approximation (cRPA) by Wehling \textit{et al.}~in Ref.~\cite{Wehling:2011df}. 
Therein exact values were obtained for the on-site $U_{00}$,
nearest-neighbor $U_{01}$, next-nearest-neighbor $U_{02}$, and
third-nearest-neighbor $U_{03}$ interaction parameters, and a
momentum dependent phenomenological dielectric screening formula
derived, based on a thin-film model, which can be used to interpolate
to an unscreened Coulomb tail at long distances. Here we use the
``partially screened Coulomb potential'' of Ref.~\cite{Smith:2014tha}
which combines both results via a parametrization based on a distance
dependent Debye mass $m_D$. The matrix elements $V_{xy}$ are then
filled using   
\begin{equation}
   V(r)  = \left\{ \begin{array}{lr}      
     U_{00},U_{01},U_{02},U_{03} & , \; r\le 2a \\[2pt]
     e^2  \left( c \, \displaystyle
     \frac{\exp(-m_D r)}{a (r/a)^{\gamma}}  + m_0
   \right)  &, \; r > 2a  
      \end{array}
          \right.\label{eq:potfit}
\end{equation}
where $a$ is the nearest-neighbor distance as before, and  $m_D$,
$m_0$, $c$ and $\gamma$ are piecewise constant chosen such that $m_D,
m_0 \to 0$ and $c, \gamma \to 1$ for $r \gg a $. For the precise
values of these parameters we refer to the tables in
\cite{Smith:2014tha}. The resulting interaction potential is shown in
comparison to the unscreened Coulomb potential in
Fig.~\ref{fig:potential}.

We note in passing that there is still some theoretical
uncertainty concerning the  screening effects generated by the
$\sigma$-band electrons at short distances (for a detailed discussion,
see Ref.~\cite{Tang:2015as}). For the purpose of our present study,
this is of minor importance because our main conclusions should be  
insensitive to small variations of the short-range interaction
parameters. Larger variations of these parameters on the other hand
can lead to very rich phase diagrams including topological insulating
phases \cite{Raghu:2008a}. A detailed study of competing order from 
HMC simulations of the extended Hubbard model on the hexagonal lattice
with varying on-site and nearest-neighbor couplings currently in
progress \cite{Buividovich:2016tgo}. 

To proceed, one derives a functional-integral formulation of the
grand-canonical partition function $Z=\Tr e^{-\beta H}$, in which the
ladder-operators are replaced by Grassman valued fermionic field
variables, by factorizing $e^{-\beta H}$ into $N_t$ terms (taken to be
``slices'' in Euclidean time) and inserting complete sets of fermionic
coherent states. Formally, $N_t$ must be taken to infinity to obtain 
an exact result, but for numerical simulations $N_t$ is a finite
number. This implies a discretization error of order
\calo($\delta^2$), where $\delta=\beta/N_t$. The final result is 

\begin{widetext}
\begin{align}
Z &= \int \prod_{t=0}^{N_t-1} \left[ \prod_x d\psi^{*}_{x,t} \, d\psi_{x,t} \, d\eta^{*}_{x,t} \, d\eta_{x,t} \right]
\exp \Big\{ -\delta  \Big[\frac{1}{2}\sum_{x,y} Q_{x,t+1,t}V_{xy}Q_{y,t+1,t}\notag\\
&-\sum_{\langle x,y \rangle}\kappa
(\psi_{x,t+1}^{*} \psi_{y,t}+\psi_{y,t+1}^{*} \psi_{x,t}+\eta_{y,t+1}^{*} \eta_{x,t}+\eta_{x,t+1}^{*} \eta_{y,t}  )
+\sum_x
  m_s(\psi_{x,t+1}^{*}\psi_{x,t}+\eta_{x,t+1}^{*}\eta_{x,t})
\notag\\
&+\frac{1}{2}
  \sum_x V_{xx}(\psi_{x,t+1}^{*}\psi_{x,t}+\eta_{x,t+1}^{*}\eta_{x,t})\Big]
-\sum_x\big[ \psi^{*}_{x,t+1} (\psi_{x,t+1}-\psi_{x,t}  )+\eta^{*}_{x,t+1} (\eta_{x,t+1}-\eta_{x,t}) \big] \Big\}~.\label{eq:partfunc1}
\end{align}
\end{widetext}
Here we have used the notation 
$Q_{x,t,t'}=\psi_{x,t}^{*}\psi_{x,t'} - \eta_{x,t}^{*}\eta_{x,t'}$.

One would now like to integrate out the fermionic fields to obtain an expression
containing only determinants of a fermionic matrix $M$, which can then be sampled stochastically. This is
prevented by fourth powers of the fields, appearing in the interaction term $\sim q_x V_{xy} q_y$. These
can be removed by a Hubbard-Stratonovich transformation 
\begin{align}
\exp\big\{& -
\frac{\delta}{2}\sum_{x,y}q_{x}V_{xy}q_{y} \big\}
\propto  \int \big[ \prod_x \phi_x \big]\, \notag\\
&\times \exp\big\{-\frac{\delta}{2}
\sum_{x,y} \phi_{x} V_{xy}^{-1} \phi_{y}
-i\,\delta
\sum_{x} \phi_{x} q_{x}\big\}~, \label{eq:hubbard1}
\end{align}
at the expense of introducing an additional dynamical scalar field $\phi$ (``Hubbard field''). The resulting expression
contains only quadratic powers, so Gaussian integration can be carried out, which yields
\begin{align}
Z =\int \big[& \prod_{x,t} \phi_{x,t} \big]\
\, \det\big[ M(\phi) M^\dagger(\phi)  \big] \notag \\
&\times\exp \big\{-\frac{\delta}{2} \sum_{t=0}^{N_t-1} \sum_{x,y}
\phi_{x,t}V_{xy}^{-1} \phi_{y,t} \big\}~, \label{eq:partfunc2}
\end{align}

A subtlety here is that, if the Hubbard-Stratonovich transformation
(\ref{eq:hubbard1}) is naively applied to Eq.~(\ref{eq:partfunc1}),
the determinant of the fermion matrix is a high-degree polynomial of
the non-compact field $\phi$ whose numerical evaluation is plagued by
uncontrollable rounding errors. It is therefore advantageous to use an
alternative fermion discretization with a coupling to a compact
Hubbard field \cite{Brower:2012zd,Ulybyshev:2013swa,Smith:2014tha}.
Its derivation is slightly more involved but straightforward,
essentially based on applying the Hubbard-Stratonovich transformation
\emph{before} introducing the fermionic coherent states. The matrix
elements are then computed using the identity
\be
\langle \xi |\, e^{\sum_{x,y}\axd A_{xy} \ay }\, |\xi' \rangle =
\exp\left( \sum_{x,y} \xi^*_x \left( e^A \right)_{xy} \xi'_y \right)~,
\label{eq:matrixelement2}
\ee
which holds for arbitrary matrices $A$. Here, $A$ is a diagonal matrix with
elements $A_{xx}=\pm i \delta\, \phi_x$. The differences are
of subleading order $\delta^2$ in the time discretization. Hence both
are equivalent at the order $\delta$ and share the same continuum
limit. It is this modified version of the fermion matrix $M(\phi)$,
with the compact Hubbard field, which is used for numerically
stablility in our simulations. Its matrix elements are given by (for
details, see Ref.~\cite{Smith:2014tha}):   
\begin{align}
&M_{(x,t)(y,t')}(\phi)=
\delta_{xy}(\delta_{tt'}-e^{-i\frac{\beta}{N_t}\phi_{x,t}} \delta_{t-1,t'})\notag \\
&\quad-\kappa\frac{\beta}{N_t} \sum\limits_{n} \delta_{y,x+\vec\delta_{n}}\delta_{t-1,t'}+ 
m_s\frac{\beta}{N_t} \delta_{xy} \delta_{t-1,t'}~.\label{eq:fermionmat1}
\end{align}
The matrix contains terms corresponding to the different contributions
from the tight-binding Hamiltonian and a covariant derivative in
Euclidean time, in which the Hubbard field enters in form of a gauge
connection where $\phi$ acts as an electrostatic potential.  
 
Both $M$ and $M^\dagger$ appear in Eq.~(\ref{eq:partfunc2}) due to the
two spin orientations entering as independent degrees of freedom into the Hamiltonian (we are essentially treating spin-up and
spin-down states as different particle flavors). The resulting
expression is suitable for simulation via HMC at half filling ($\mu =
0$), as the integrand may be interpreted as a weight function for 
the Hubbard field $\phi$. 
 
\subsection{Hybrid-Monte-Carlo and the fermion-sign problem}
\label{sec:HMCfsp}
The HMC method (originally developed for strongly interacting
fermionic quantum field theories \cite{Duane:1987de})
consists in essence of creating a distribution
of field configurations representative of the thermal equilibrium, by evolving the $\phi$ field in
computer time $\tau$ through a fictitious deterministic dynamical process, governed by a conserved
classical Hamiltonian defined in the higher dimensional space
spanned by real Euclidean space-time and $\tau$. Quantum fluctuations enter in the form of stochastic
refreshments of the canonical momentum $\pi$ associated with the Hubbard field $\phi$. 
As a symplectic integrator must be used to solve Hamilton's equations for $\phi$ and $\pi$, an additional
error arises from the finite step-size of this integrator, which is subsequently corrected by a Metropolis
accept/reject step. HMC is thus an exact algorithm (see Ref.~\cite{Smith:2014tha} for further details).

HMC is a form of \emph{importance sampling}, i.e.~a method of approximating the functional integral by 
probabilistically generating points in configuration space which are clustered in the regions that 
contribute most to the integral. A crucial criterion for its applicability is the existence of a
real and positive-definite
measure for the dynamical fields, which may then be interpreted as a probability density. 
This is true here only
because the phases of $M$ and $M^\dagger$ cancel exactly in
Eq.~(\ref{eq:partfunc2}).  
As we will see, this no longer holds at non-zero charge density. 

To generate finite charge-carrier density, one would have to add a
corresponding chemical potential $\mu$, replacing the
Hamiltonian in Eq.~(\ref{eq:tightbinding}) by
\begin{equation}
H \to H - \mu \sum_x q_x  = H - \mu \sum_x(\axd \ax - \bxd \bx )\, . 
\label{eq:tightbinding_withmu}
\end{equation}
At the level of the partition function, this leads to the modification
\begin{align}
&Z(\mu) = \int \prod_{t=0}^{N_t-1} \left[ \prod_x d\psi^{*}_{x,t} \, d\psi_{x,t} \, d\eta^{*}_{x,t} \, d\eta_{x,t} \right]
\notag \\&~\times
\exp \Big\{ \left( \ldots \right) + \frac{\beta\mu}{N_t} \sum_x
(\psi_{x,t+1}^{*} \psi_{x,t}-\eta_{x,t+1}^{*} \eta_{x,t}) \Big\}~.\label{eq:partfunc3}
\end{align}
After integrating out the fermion fields, one obtains a modified version of Eq.~(\ref{eq:partfunc2})
\begin{align}
Z =&\int \big[ \prod_{x,t} \phi_{x,t} \big]\
\, \det\big[ M(\phi,\mu) \widetilde{M}(\phi,\mu)  \big]\notag\\
&\times \exp \big\{-\frac{\delta}{2} \sum_{t=0}^{N_t-1} \sum_{x,y}
\phi_{x,t}V_{xy}^{-1} \phi_{y,t} \big\}~, \label{eq:partfunc4}
\end{align}
where
\begin{align}
M(\phi,\mu)_{(x,t)(y,t')}=& M(\phi,0)_{(x,t)(y,t')}
 - \mu \frac{\beta}{N_t} \delta_{xy} \delta_{t-1,t'}~,\notag\\
\widetilde{M}(\phi,\mu) _{(x,t)(y,t')}=& M^\dagger(\phi,0)_{(x,t)(y,t')}
 + \mu \frac{\beta}{N_t} \delta_{xy} \delta_{t-1,t'} \notag\\
=& M^\dagger(\phi,-\mu)_{(x,t)(y,t')}
  ~.
\label{eq:fermionoperators}
\end{align}
There is no cancellation of phases in Eq.~(\ref{eq:partfunc4}), thus importance sampling breaks
down, as we no longer can interpret the integrand as the weight of a given microstate 
in the ensemble. This is at the root of the fermion-sign problem. Whether it is a hard problem or not depends on the expectation value of the phase of the determinant in the ``phase-quenched'' theory defined by the modulus of the fermion determinant in the measure, i.e. writing
\begin{align}
Z =&\int \big[ \prod_{x,t} \phi_{x,t} \big]\
\, \big|\det M(\phi,\mu) \big|^2 \, \frac{\det\widetilde{M}(\phi,\mu)}{\det\widetilde{M}(\phi,-\mu)} \notag\\
&\times \exp \big\{-\frac{\delta}{2} \sum_{t=0}^{N_t-1} \sum_{x,y}
\phi_{x,t}V_{xy}^{-1} \phi_{y,t} \big\}~, \label{eq:partfunc4a}
\end{align}
we consider the complex ratio of determinants with opposite-sign chemical potentials as an observable in the phase-quenched theory with partition function $Z_\mathrm{pq}$  and
\begin{equation}
  \frac{Z(\mu)}{Z_\mathrm{pq}(\mu)} \, = \, \Big\langle
  \frac{\det\widetilde{M}(\phi,\mu)}{\det\widetilde{M}(\phi,-\mu)}
  \Big\rangle_\mathrm{pq} \, . \label{eq:reweight}
\end{equation}
Obviously this ratio is unity at half filling (i.e.~for $\mu\to 0$) and at
vanishing interaction strength  for all $\mu$, because the non-interacting
tight-binding theory is blind to the sign of $\mu$ for each spin
component individually. 

\begin{figure}
\begin{center}
\includegraphics[width=0.97\linewidth]{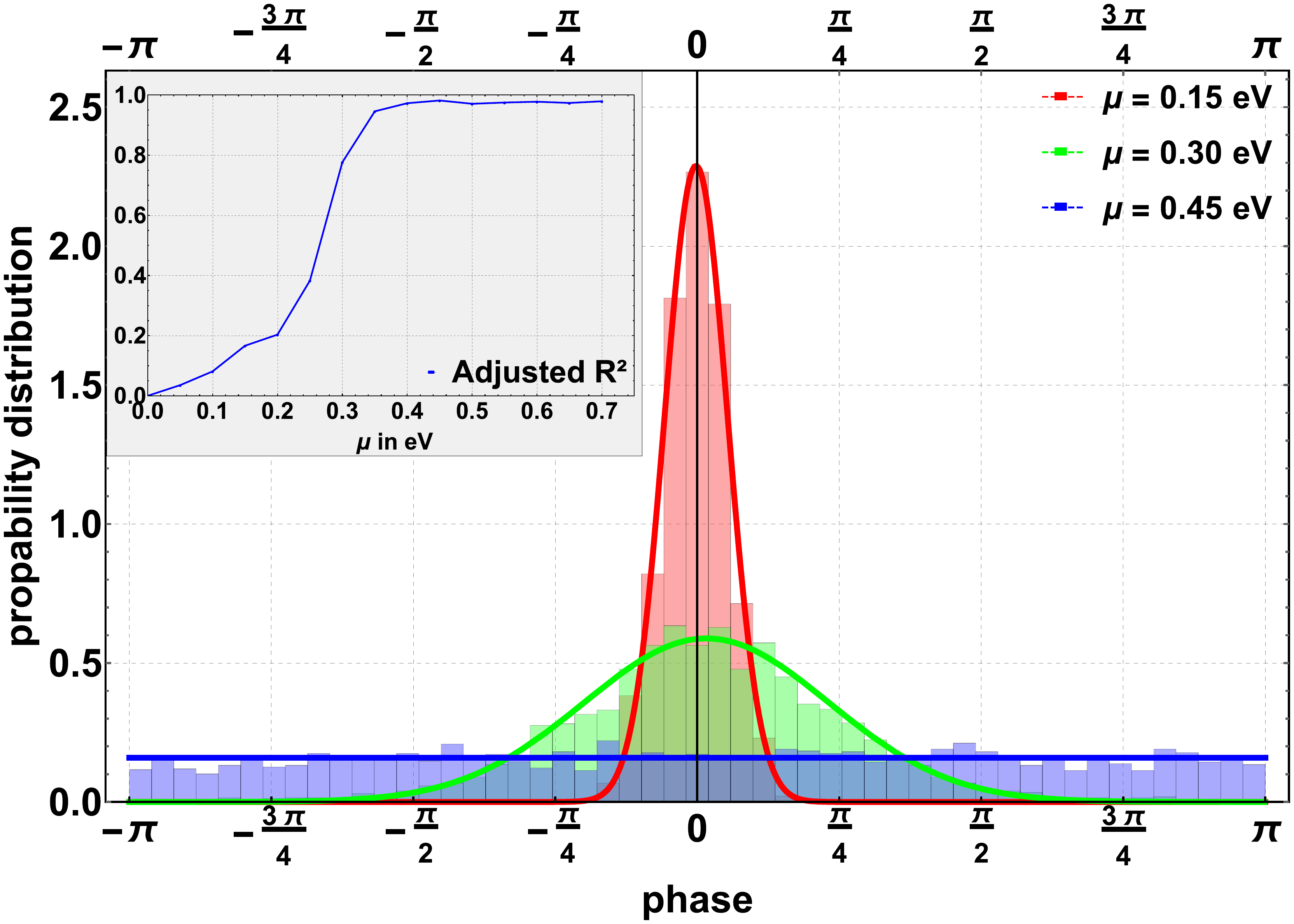}
\caption{Histograms of the phase of
  $\det\widetilde{M}(\phi,\mu)/\det\widetilde{M}(\phi,-\mu)$ obtained
  from a $6\times 6 $ lattice at $\beta = 2 \textrm{ eV}^{-1}$ for
  different $\mu$, at $10\, \%$ of the interaction strength of
  suspended graphene. The results are modelled with gaussian ($\mu =
  0.15\textrm{ eV}^{-1}$ and $0.30\textrm{ eV}^{-1}$) and uniform
  ($\mu = 0.45\textrm{ eV}^{-1}$) distributions respectively. The
  inlay shows the adjusted $\textrm{R}^2$ for fitting a constant to
  the data at a range of different $\mu$. For $\mu \gtrsim 0.4\textrm{
    eV}$ the numerical datais well described by a uniform
  distribution, indicating a hard sign problem.} 
\label{fig:signproblem}
\end{center}
\end{figure}

To exemplify that the signal is indeed lost quickly, however, when the
chemical potential for charge-carrier density is tuned away from half
filling in the interacting theory, we have measured the modulus and
the complex phase of the ratio of determinants in
Eq.~(\ref{eq:reweight}) on a $6 \times 6$
  lattice, at $\beta=2\, \textrm{eV}^{-1}$ and $10\, \%$ of the
  interaction strength of suspended graphene. 
This method of ``reweighting'' therefore certainly fails near the van Hove
singularity, already at rather moderate interaction
strengths. Fig.~\ref{fig:signproblem} shows histograms of the 
phase for different values of $\mu$ together with fit-model curves. 
As a measure for the signal-to-noise ratio we have used the adjusted
$R^2$ associated with attempting to model the histograms with a uniform distribution (this quantity is $0$ for a strictly non-linear relation between
the data and the fitted curve and $1$ for a perfect linear dependence). 
As one can see in the figure, the adjusted $R^2$ of the constant fit
shows a rather rapid crossover and approaches values close to $1$ at $\mu
\approx 0.4 \textrm{eV}$ which indicates that the signal is lost in
the noise already on the $6\times 6$ lattice. The effect will be
further enhanced with increasing lattice sizes. Note that the
modulus of the ratio of determinants is not unity here eihter. In
fact, it also decreases with $\mu$. As usual, however, it is the phase
fluctuations that are primarily responsible for the loss of signal due
to cancellations. 

The underlying physical reason for a non-polynomially hard
signal-to-noise-ratio problem typically is that the overlap of
phase-quenched and full ensembles tends to zero exponentially
because of a complete decoupling of the corresponding Hilbert spaces in
the infinite-volume limit when the two ensembles correspond to
excitations above  different finite-density ground states (here
charge-carrier versus spin density). An exponential error reduction
might be possible with generalized density-of-states methods
\cite{Langfeld:2012ah} which work beautifully in spin systems
\cite{Langfeld:2014nta} and heavy-dense QCD \cite{Garron:2016noc} but
have yet to be applied to strongly interacting theories with dynamical
fermions.  

Dense fermionic theories with a sign problem are a very active field
of research and we cannot cover the vast body of literature
here. There is no general solution, however. 

Sometimes cluster algorithms \cite{Chandrasekharan:1999cm} or
extensions thereof that exploit cancellations of field configurations
\cite{Huffman:2013mla} help.
On the other hand, when they do, there also appears to be  
an underlying Majorana positivity \cite{Li:2015a,Wei:2016a,Li:2016a}
and the theory therefore really is sign-problem free as in the case of
the anti-unitary symmetries such as time-reversal invariance with
Kramers degeneracy discussed below.

Sometimes it is possible to simulate dual
theories with worm algorithms \cite{Prokofev:2001ddj,Mercado:2013ola}. 
Deformation of the originally real configuration space into a
complex domain can help by either sampling Lefschetz
thimbles of constant phase \cite{Mukherjee:2014hsa}, reducing the sign
problem to that of the residual phases, or more generally, field
manifolds with a milder sign problem obtained from holomorphic gradient flow 
\cite{Alexandru:2016ejd}. Doubling the number of degrees of freedom by
complexification one can also try a complex version of
stochastic quantization, i.e.~by simulating the corresponding Complex
Langevin process \cite{Sexty:2014dxa}.  

While all these techniques have their difficulties and are actively
being further developed, in the mean time we follow a different
strategy here. This is to simulate a sign-problem free variant of the
original theory with standard Monte-Carlo techniques and study genuine
finite-density effects where importance sampling is possible. Such
variants could be theories with anti-unitary symmetries such as
two-color QCD, with two instead of the usual three colors
\cite{Boz:2015ppa,Holicki:2017psk}, or 
 $G_2$-QCD, with the exceptional Lie group $G_2$ replacing the $SU(3)$
gauge group of QCD \cite{Maas:2012wr,Wellegehausen:2015iea}.

The arguably simplest variant is the phase-quenched theory itself,
however. In two-flavor QCD this amounts to simulating at finite
isospin density \cite{Detmold:2012wc,Brandt:2016zdy}. Here it
corresponds to introducing a chemical potential for finite spin 
density, like a pure Zeeman term from an in-plane magnetic field, rather
than one for finite charge-carrier density, as mentioned above. 
To this end we add a chemical potential $\mu_\sigma = (-1)^\sigma
\mu $ with a spin $\sigma = 0,1$ (for up/down) dependent sign,
i.e.~instead of (\ref{eq:tightbinding_withmu}) we use the replacement
\begin{equation}
H \to  H -   \mu  \sum_{x}  (\axd \ax + \bxd \bx )~.
\label{eq:tightbinding_withisomu}
\end{equation}
Compared to (\ref{eq:tightbinding_withmu}), the sign of the
term $\sim\bxd \bx$ has been flipped. This leads to a modification of
the spin-down determinant in Eq.~(\ref{eq:fermionoperators}), such that
\begin{align}
\widetilde{M}(\phi,\mu_\sigma) _{(x,t)(y,t')}=& M^\dagger(\phi,0)_{(x,t)(y,t')}
 - \mu \frac{\beta}{N_t} \delta_{xy} \delta_{t-1,t'}\notag\\
 =& M^\dagger(\phi,\mu )_{(x,t)(y,t')}~.
\end{align}
Cancellation of the phases in the partition function is thus restored;
$\mu_\sigma$ shifts the Fermi surfaces for electron-like and hole-like
excitations in opposite directions. As the nearest-neighbor
tight-banding bands are symmetric under exchange of particle-like and
hole-like states individually for each spin, the Lifshitz transition
in the non-interacting theory is in fact blind to this change of
sign. As a result, $\mu_\sigma$ induces a Zeeman-splitting but without the phase
factors from a Peierls substitution in the hopping
term. It therefore describes graphene coupled to an in-plane magnetic
field \cite{Aleiner:2007va}. In the following we will omit the
spin-index. It is implied that $\mu $ is spin staggered from now on,
i.e.~corresponding to $\mu_\sigma = (-1)^\sigma \mu $ as in
Eq.~(\ref{eq:tightbinding_withisomu}). 

\subsection{Observables}
\label{subsec:Observables}
Expectation values of physical operators in the thermal ensemble are
expressed in the path-integral formalism as
\eq{\langle O \rangle=\frac{1}{Z} \int D\phi\, O(\phi)\, 
\text{det}\left(MM^\dag \right) e^{-S(\phi)}~. }
Their representation in the space of field variables can be obtained
from derivatives of the partition function with respect to
corresponding source terms. We are interested in the 
particle-hole susceptibility (\ref{eq:TFs}), which up to a factor
of $\beta=1/T$ agrees with the number susceptibility (per unit
cell).\footnote{Of course, with the spin-staggered $\mu$ it is strictly
speaking not a number but a spin, i.e.~magnetic susceptibility, see above.} 
Hence it is given by 
\eqa{ \chi(\mu)  &= - \frac{1}{N_c} \,
  \left(\frac{d^2\Phi}{d\mu^2} \right) 
\\&= \frac{1}{N_c \beta} \left[\frac{1}{Z} \frac{d^2Z}{d\mu^2}
  -\frac{1}{Z^2} \left(\frac{dZ}{d\mu}\right)^2 \right]~, \notag}
where  $\Phi=-T \ln{Z}$ is the grand-canonical potential and $N_c=N^2$ is the
number of unit cells.  Using the path-integral representation of $Z$,
we can express  $\chi(\mu) $ in terms of the fermion
matrix $M(\phi)$, since 
\eq{\frac{1}{Z} \frac{d^n Z}{d\mu^n} = \frac{1}{Z} \int D\phi \left[ \frac{d^n}{d\mu^n} \text{det}\left(MM^\dag \right) \right] e^{-S(\phi)}~. }
Calculating the derivatives for $n=1,2$ we obtain 
\eq{\hspace{-4mm}\frac{d}{d\mu} \text{det}\left(MM^\dag\right) = 2\, \text{det}\,\left(MM^\dag\right)\,
\text{ReTr}\,\left(M^{-1} \frac{dM}{d\mu}\right)}
and
\begin{align}
\hspace{-1mm}\frac{d^2}{d\mu^2} \text{det}&\left(MM^\dag\right) = 4\, \text{det}\left(MM^\dag\right) \left\{ \left[ \text{ReTr}\,\left( M^{-1} \frac{dM}{d\mu} \right) \right]^2 \right.\notag\\
 &\left.\quad\quad\quad - \frac{1}{2} \, \text{ReTr}\,\left( M^{-1} \frac{dM}{d\mu}M^{-1} \frac{dM}{d\mu} \right) \right\}.
\end{align}

\noindent Using these relations we can write the spin-staggered particle-hole
susceptibility as  
$\chi = \chi_\textrm{con} + \chi_\textrm{dis} $, with
\begin{align}
\chi_\textrm{con}(\mu)  =& \frac{-2}{N_c \beta}
\left\langle  \text{ReTr}\,\left( M^{-1} \frac{dM}{d\mu}M^{-1} \frac{dM}{d\mu} \right) \right\rangle\notag\\
\chi_\textrm{dis}(\mu)  =& \frac{4}{N_c \beta}\left\{ \left\langle \left[ \text{ReTr}\,\left( M^{-1} \frac{dM}{d\mu} \right) \right]^2
\right\rangle \right.\notag \\& \left. - \left\langle \text{ReTr}\,\left( M^{-1} \frac{dM}{d\mu} \right) \right\rangle^2 \right\}~,
\label{eq:susc1}
\end{align}
where $\chi_\textrm{con/dis}$ denote the connected and disconnected
contributions respectively. 
The brackets on the right-hand sides of Eqs.~(\ref{eq:susc1}) are understood
as averages over a representative set of field configurations. The
traces can be evaluated with noisy estimators. 

A susceptibility $\chi^\textrm{sdw}$ corresponding to the
fluctuations of the antiferromagnetic spin-density 
wave order parameter computed at half filling in
Ref.~\cite{Buividovich:2016tgo} 
can be obtained in complete analogy to the above,
replacing all derivatives with respect 
to $\mu$ by derivatives with respect to the sublattice-staggered mass
$m_s = (-1)^s \, m$  in Eq.~(\ref{eq:tightbinding}). 
The resulting expressions are then of precisely the same form as
Eqs.~(\ref{eq:susc1}), with the replacement $\mu \to m $,
\begin{align}
\chi_\textrm{con}^\textrm{sdw}(\mu)  =& \frac{-2}{N_c \beta}
\left\langle  \text{ReTr}\,\left( M^{-1} \frac{dM}{dm}M^{-1} \frac{dM}{dm} \right) \right\rangle\notag\\
\chi_\textrm{dis}^\textrm{sdw}(\mu) =& \frac{4}{N_c \beta}\left\{ \left\langle \left[ \text{ReTr}\,\left( M^{-1} \frac{dM}{dm} \right) \right]^2
\right\rangle \right.\notag \\& \left. - \left\langle \text{ReTr}\,\left( M^{-1} \frac{dM}{dm} \right) \right\rangle^2 \right\}~.
\label{eq:sdwsusc}
\end{align}

\section{Results}
\label{sec:Results}
In this chapter we first present our results for the susceptibility
$\chi(\mu)$ of ferromagnetic spin-density fluctuations, i.e.~the
spin-staggered particle-hole susceptibility, from Hybrid-Monte-Carlo
simulations of the interacting tight-binding theory at finite
spin-density and temperature. 
Only in the last subsection we briefly come back to the spin-density
dependence of the antiferromagnetic SDW susceptibility
$\chi^\textrm{sdw}$ as well. 

All results were obtained from hexagonal lattices of finite size with 
periodic Born-von K\'arm\'an boundary conditions, with an equal number of unit
cells in each principal direction. We chose a sublattice and
spin-staggered mass $m_s$ of magnitude $m = 0.5 \textrm{ eV}$, an
inter-atomic spacing of $a = 1.42 \text{ \AA}$  
and a hopping parameter of $\kappa=2.7\textrm{ eV}$.
We furthermore use the partially screened Coulomb potential
discussed in detail in Section \ref{sec:Setup}
and Ref.~\cite{Smith:2014tha}.

The rescaled effective interaction strength $\alpha_\textrm{eff}$ is defined 
in the following as $\alpha_\textrm{eff} = \lambda  \cdot
\alpha_\textrm{graphene}$ with $\alpha_\textrm{graphene} =
\frac{e^2}{\hbar v_F} \approx 2.2$ ($\lambda$ thus acts as a global
rescaling factor which changes each element of the interaction matrix
in the same way, i.e. $V_{xy} \to \lambda V_{xy}$). 
Interactions were rescaled to different magnitudes in the
range $\lambda=[0,1]$ (spanning the range from no interactions
to suspended graphene, i.e.~without any substrate induced dielectric
screening). 

For each set of parameters presented in the following, measurements were done
in thermal equilibrium on at least 300 independent configurations of
the Hubbard field. 
Integrator stepsizes were tuned such that the Metropolis acceptance rate was
always above $70\%$. All error bars were calculated taking possible
autocorrelations into account, using the binning method and standard
error propagation where appropriate. For calculation of observables
all traces are estimated with 500 gaussian noise vectors. 

\subsection{Influence of the Euclidean-time discretization}\label{sec:DISK}
As HMC simulations are carried out at finite discretization $\delta$
of the Euclidean time axis (which is related to the temperature
through the relation $\beta=\delta N_t$, where $N_t$ is the number
of time-slices), exact quantitative results can be only obtained
by $\delta \to 0$ extrapolation. As it would be  computationally prohibitively 
expensive to simulate for a suitable range of $\delta$ values with each set
of physical parameters (in particular when temperatures are low, system sizes
are large or interactions are strong) we carry out such an 
extrapolation only for a few exemplary cases. This will help to develop
an understanding of the systematics of the discretization errors
in order to assess whether simulations with a fixed discretization
can provide reliable results at reasonable cost, 
in particular for the low temperatures which are required to detect deviations
from the logarithmic divergence of $\chi(\mu=\kappa)$. Such is the
purpose of this section.  

\begin{figure}
\begin{center}
\includegraphics[width=0.97\linewidth]{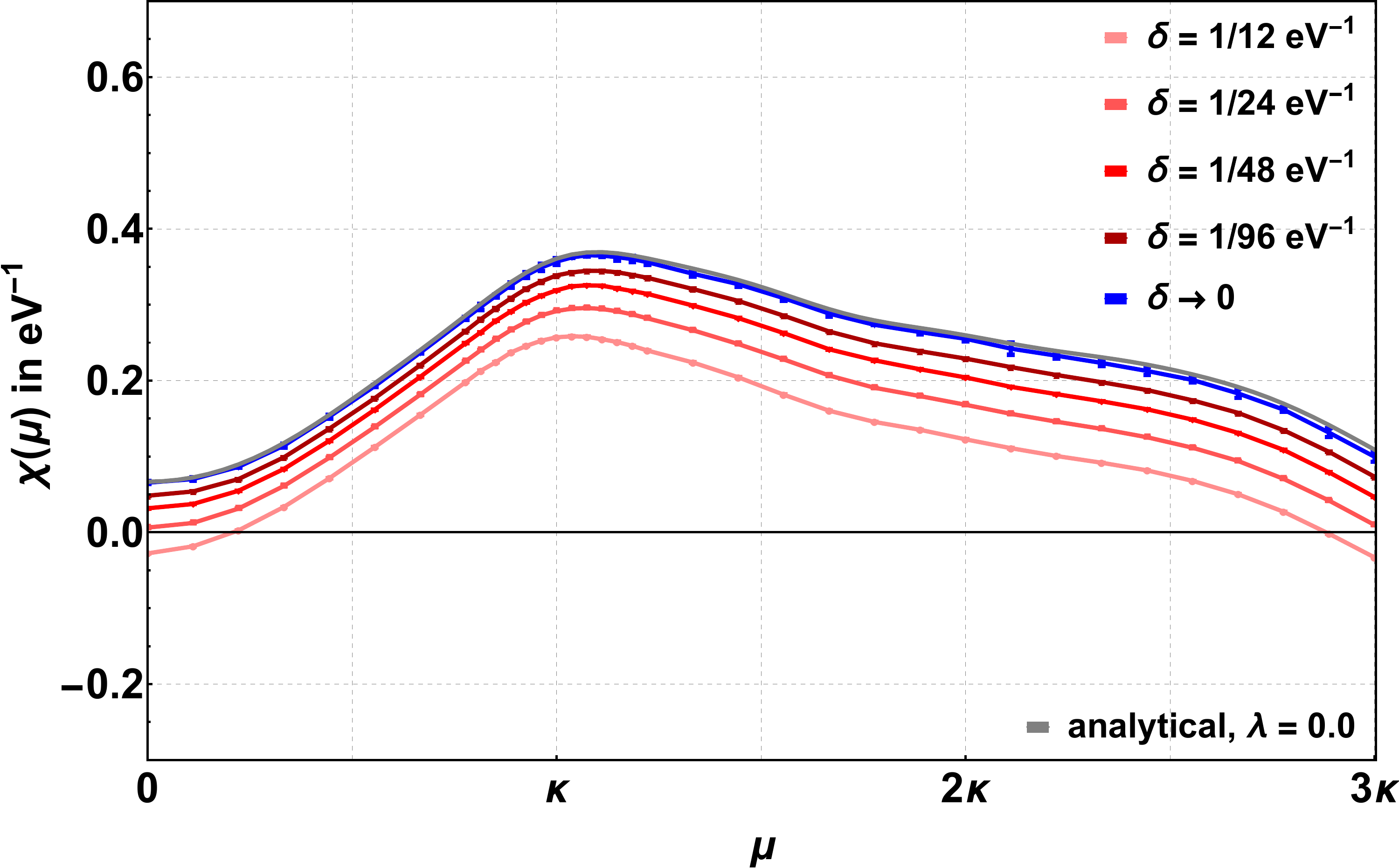}\\
\includegraphics[width=0.97\linewidth]{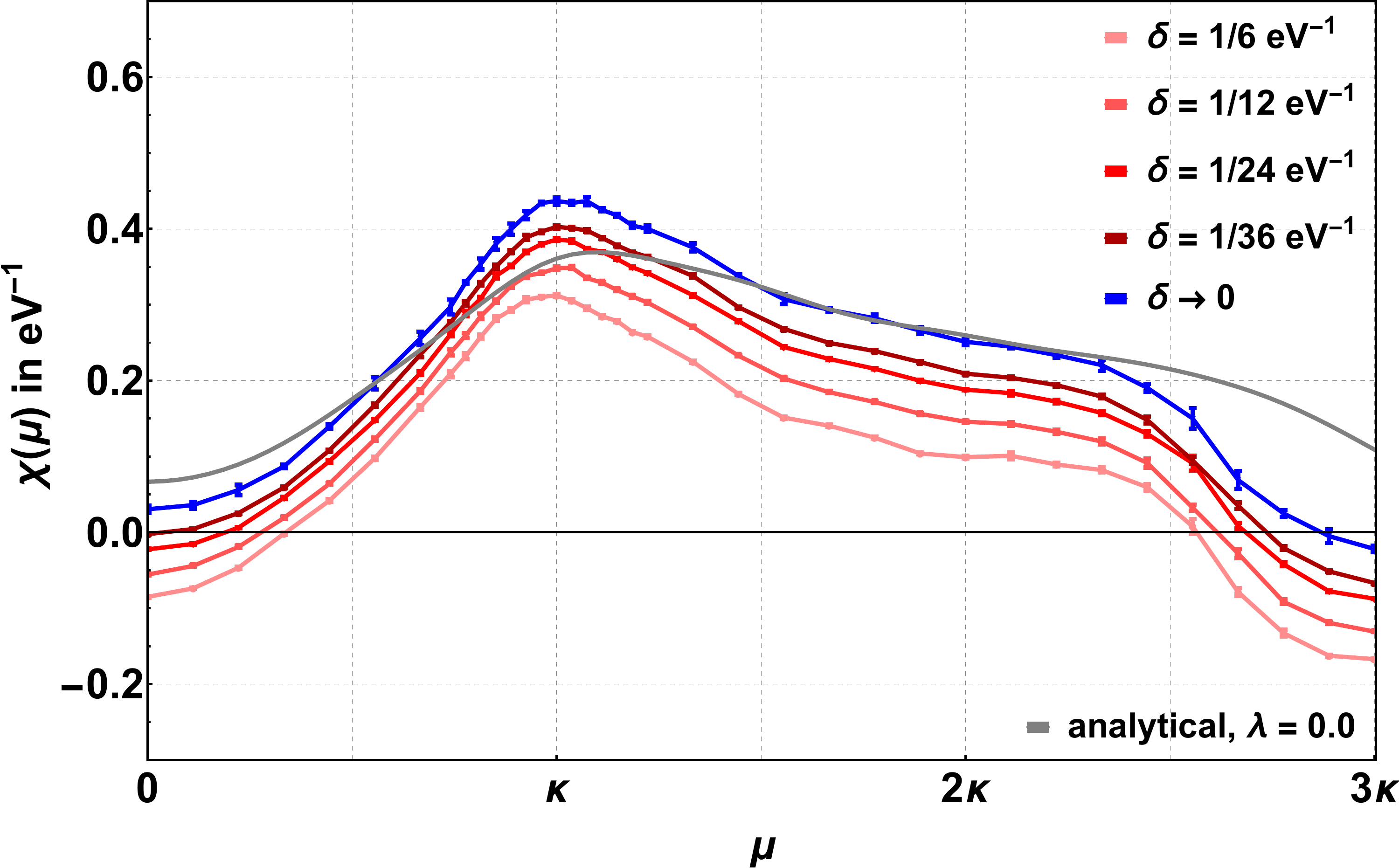}
\includegraphics[width=0.97\linewidth]{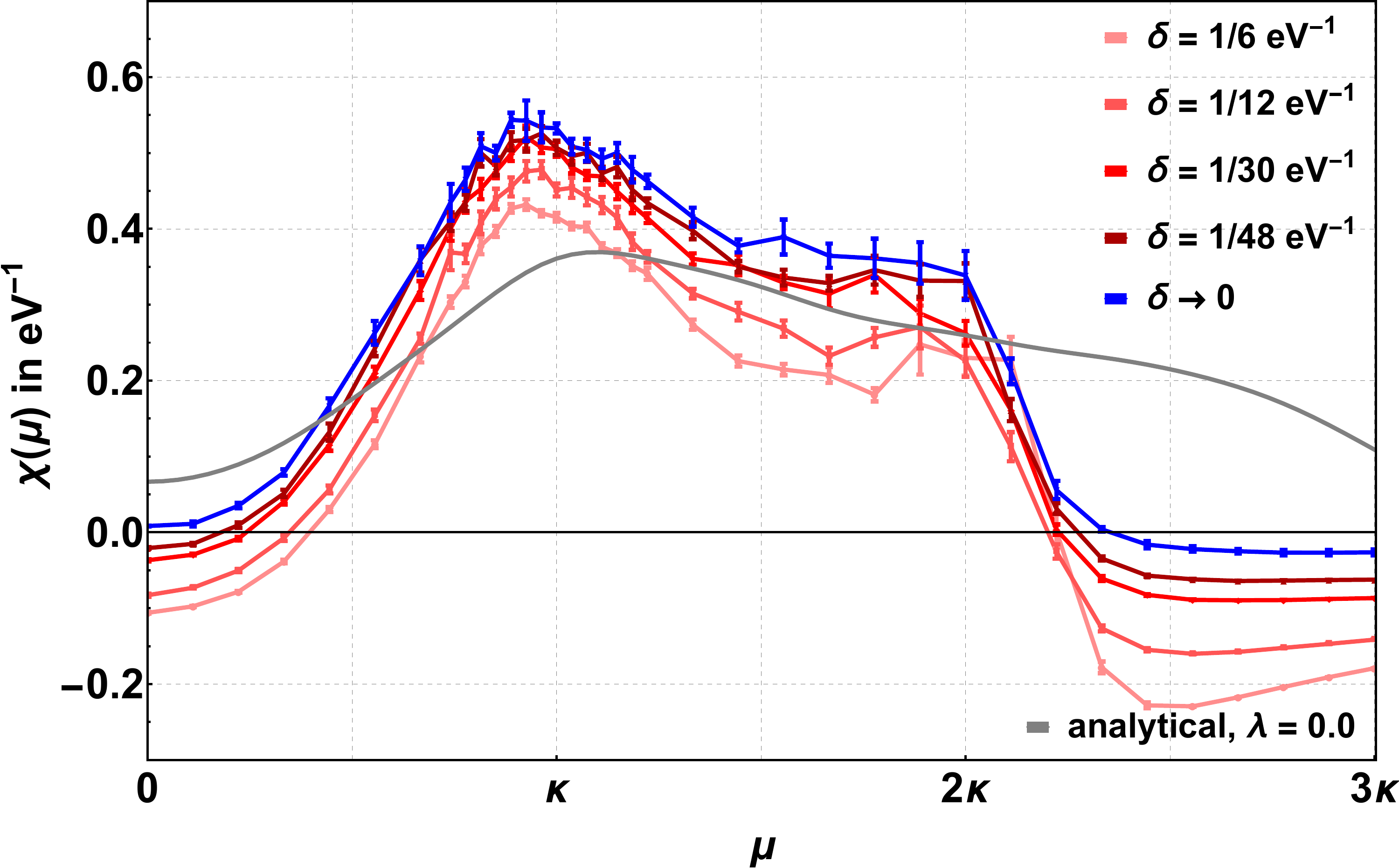}
\caption{$\chi(\mu)$ for $\lambda=0.0$ (top), $\lambda=0.4$ (middle) and $\lambda=1.0$ (bottom)
at $\beta = 2\textrm{ eV}^{-1}$, $N =12$. Different discretizations
are shown as well as pointwise quadratic $\delta \rightarrow 0$ extrapolations (blue). The semi-analytic 
$\lambda=0.0$ result obtained from Eq.~(\ref{eq:chisum}) is shown for comparison in all plots (gray).}
\label{fig:ALP_DISK}
\end{center}
\end{figure} 

\begin{figure}
\begin{center}
\includegraphics[width=0.97\linewidth]{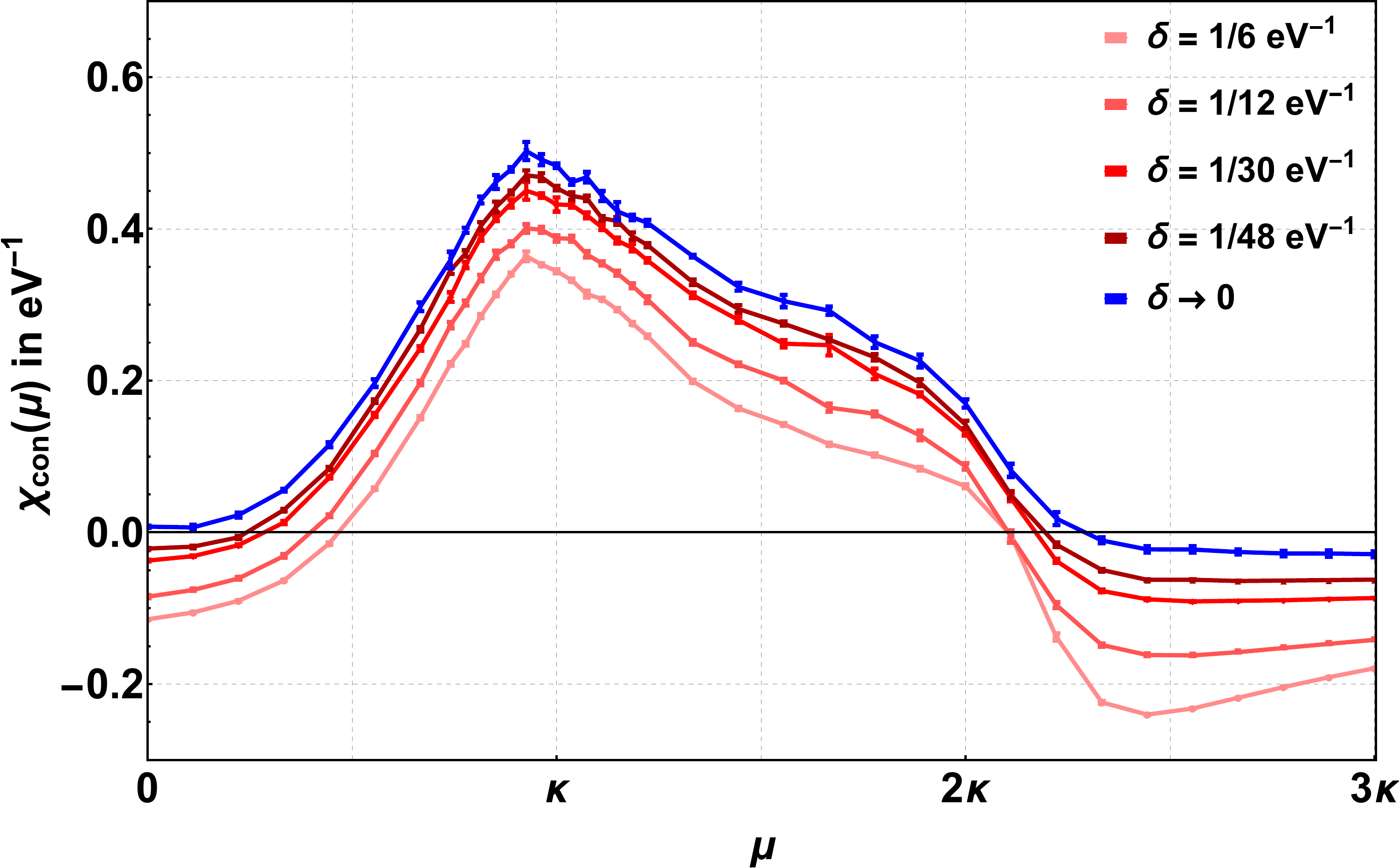}
\includegraphics[width=0.97\linewidth]{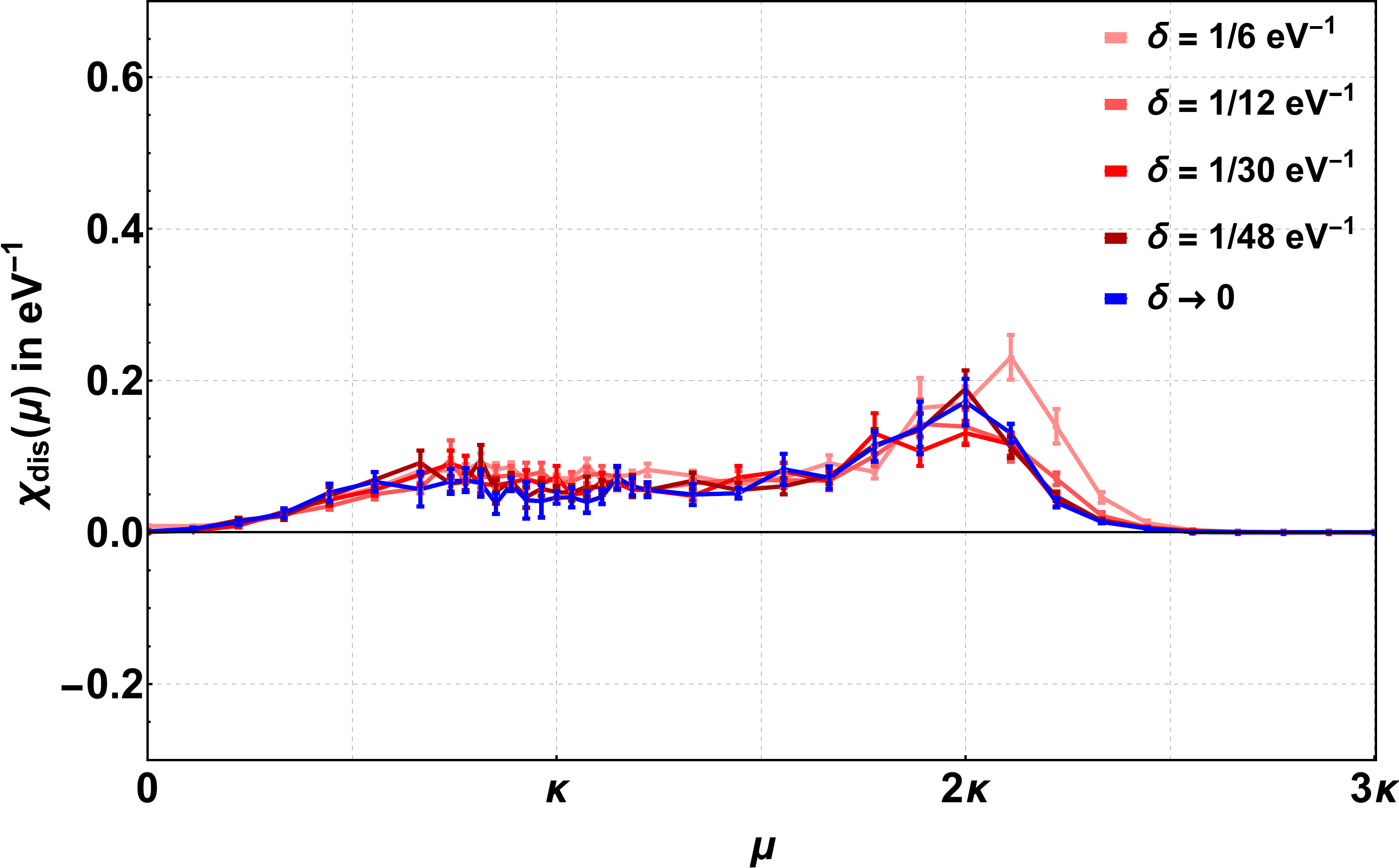}
\caption{$\chi_\textrm{con}(\mu)$ (top) and $\chi_\textrm{dis}(\mu)$ (bottom) for $\lambda=1.0$ at 
$\beta = 2\textrm{ eV}^{-1}$, $N =12$ for different discretizations (red) and 
their pointwise quadratic $\delta \rightarrow 0$ extrapolations (blue). }
\label{fig:ALP_A100_DISK_CON_DIS}
\end{center}
\end{figure}

Fig. \ref{fig:ALP_DISK} (top) shows the trivial case of $\chi(\mu)$ 
at vanishing two-body interactions, corresponding to a Hubbard field $\phi$ 
which is set to zero on all lattice sites. The inversions of the fermion matrix 
in Eqs.~(\ref{eq:susc1}) are straightforward to carry out in this case
and no molecular dynamics
trajectories are in fact needed at all. Furthermore,
the disconnected part of $\chi(\mu)$ vanishes exactly in this case, as
the expectation value $\langle \ReTr(\ldots)^2\rangle$ factorizes.
The different curves represent calculations for different values of
$\delta$, on an $N=12$ 
lattice at $\beta=2\textrm{ eV}^{-1}$ (from Fig.~\ref{DIFF_TEMP_OVER_VOL} 
we know that finite-size effects can be neglected for this choice), 
together with a point-by-point $\delta\to 0$ extrapolation using
quadratic polynomials. 
As we expect, the extrapolated points agree well with the semi-analytic
calculation from Eq.~(\ref{eq:chisum}), with small deviations only arising
from the uncertainty associated with the fitting procedure.   
We also see that the main effect of finite $\delta$ is a shift to lower 
and in some areas negative values. Fortunately, the shift is nearly
constant over the entire range of $\mu$.  
A similar behaviour can be seen when interactions are switched on.  
Figs.~\ref{fig:ALP_DISK} (middle and bottom) again show results
from the $N=12$ lattice at 
$\beta=2\textrm{ eV}^{-1}$ (for $N_t$ between 12 and 96)
but with non-zero interaction strengths corresponding to $\lambda=0.4$
and $\lambda=1.0$ respectively. For comparison, as a first indication
of the effects of interactions, we also 
show the non-interacting limit in these figures. In order to
illustrate the origin of the  discretization errors in the interacting case, in Figs.~\ref{fig:ALP_A100_DISK_CON_DIS} we also
display $\chi_\textrm{con}(\mu)$ (top) and $\chi_\textrm{dis}(\mu)$ separately for $\lambda=1.0$. What is striking is that
the disconnected part seems to depend only very weakly on $\delta$,
while the connected part displays the familiar shift. This is a
fortunate situation, as it is $\chi_\textrm{dis}$ which is expected to
show the characteristic scaling indicative of a true thermodynamic
phase transition. 

Our main conclusion here is that we have good justification to assume
that the effect of interactions can be studied qualitatively rather
well for fixed $\delta$. Nevertheless, we present a set of fully
extrapolated results for $\beta=2\textrm{ eV}^{-1}$ in the following
section. Results for lower temperatures will then be presented for
fixed $\delta$. 

\subsection{Influence of inter-electron interactions}\label{sec:manybody}

\begin{figure}
\begin{center}
\includegraphics[width=0.97\linewidth]{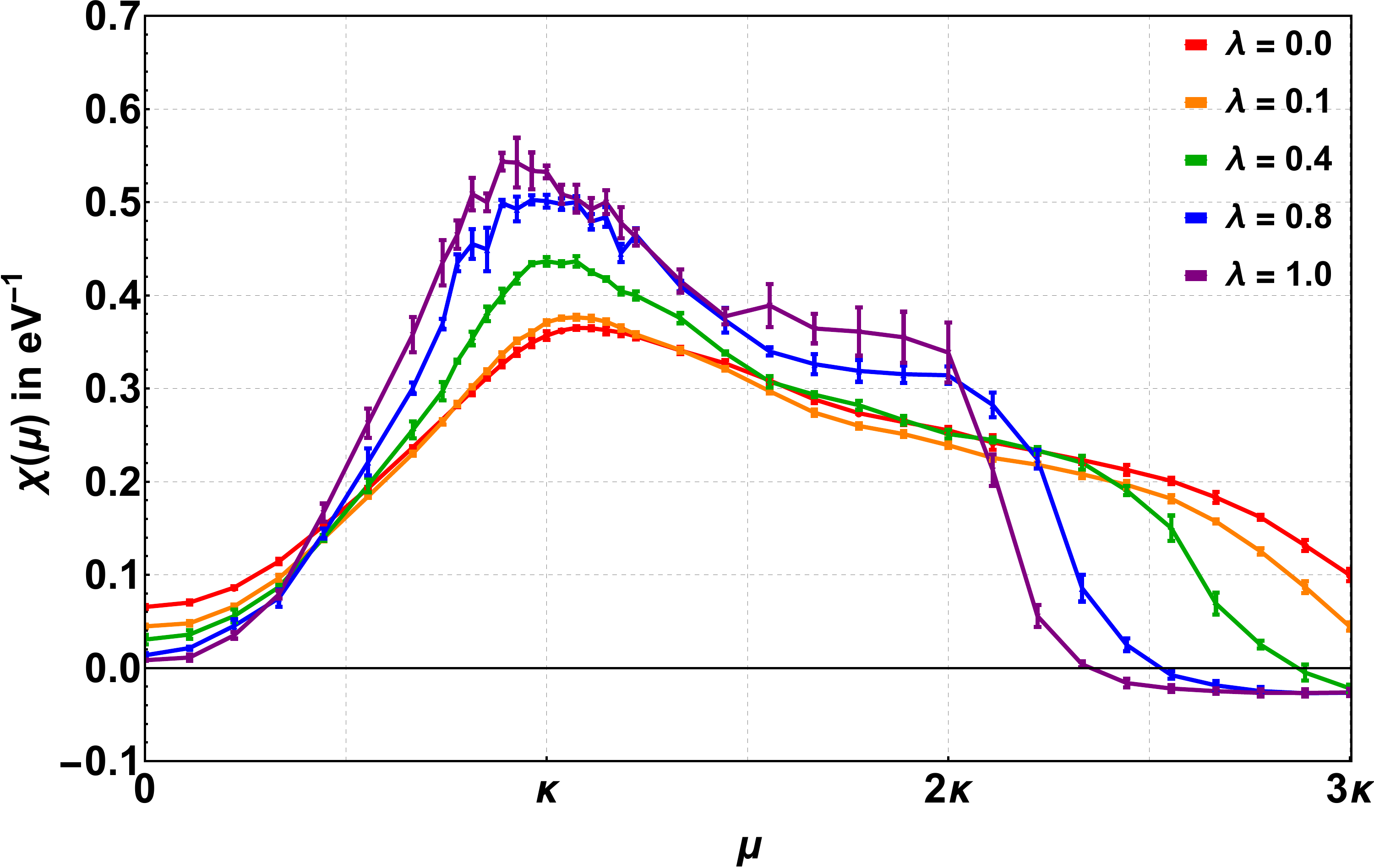}
\includegraphics[width=0.97\linewidth]{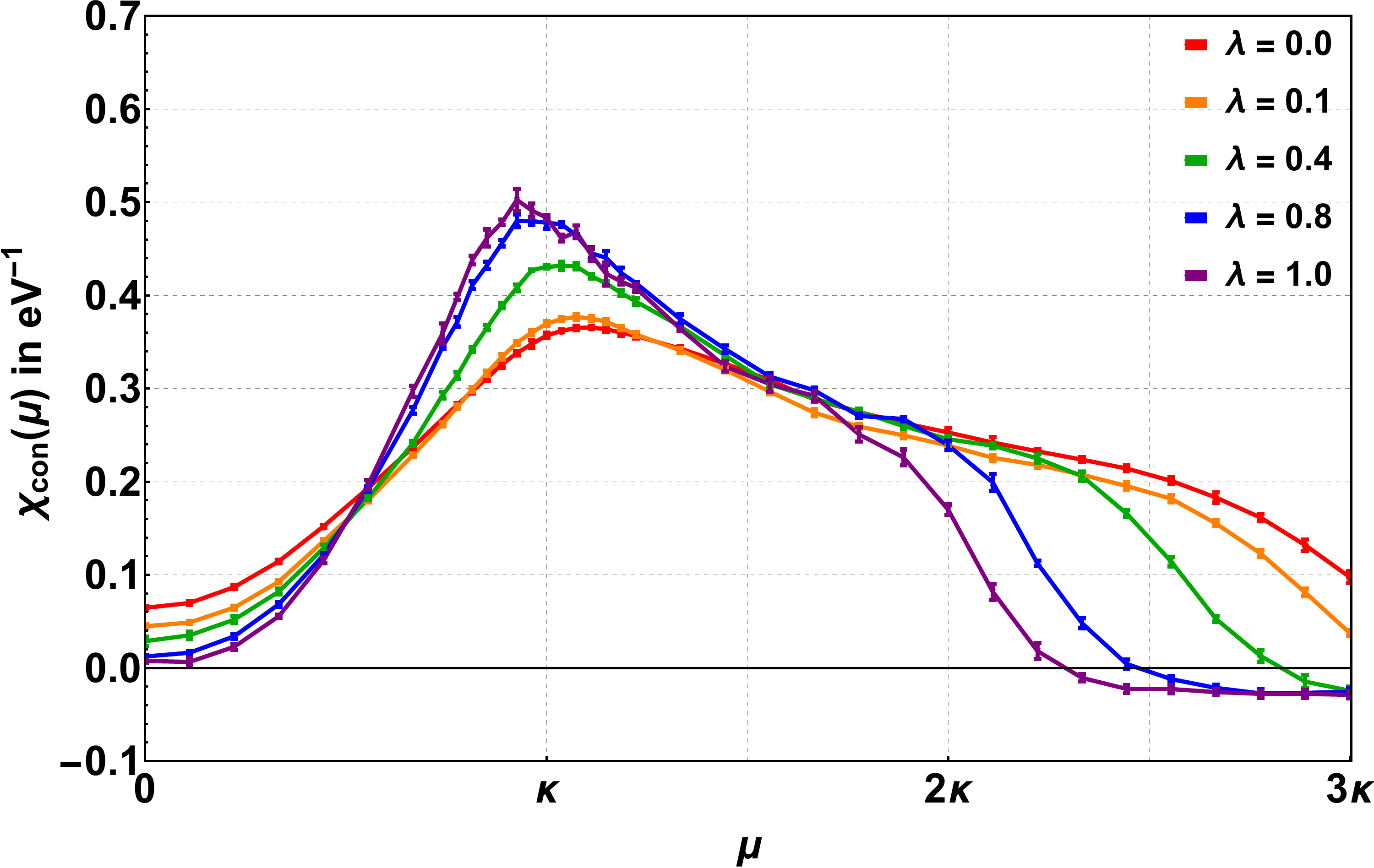}
\includegraphics[width=0.97\linewidth]{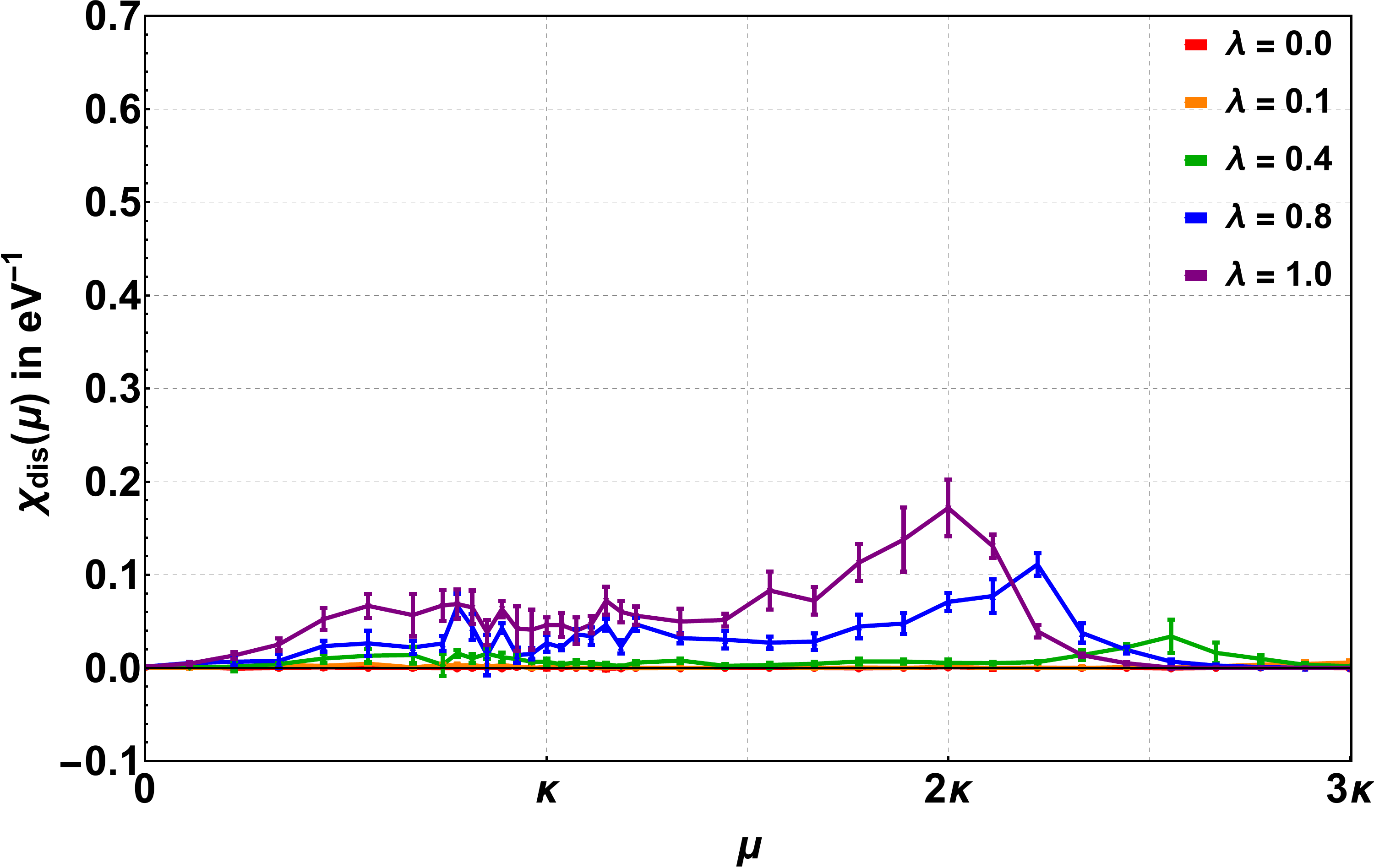}
\caption{$\chi(\mu)$ (top), $\chi_\textrm{con}(\mu)$ (middle) and $\chi_\textrm{dis}(\mu)$ (bottom) 
for $\beta=2\textrm{ eV}^{-1}$, $N = 12$ at different interaction strengths. 
All displayed points are quadratic $\delta \rightarrow 0$ extrapolations
from simulations at non-zero $\delta$. }
\label{fig:susB2V12M5}
\end{center}
\end{figure}

To demonstrate the effects of inter-electron interactions we have
carried out the same $\delta \to 0$
extrapolations for 
$\beta = 2\textrm{ eV}^{-1}$ , $N = 12$, and $\lambda \in
\left\{0.1,0.4,0.8,1.0 \right\}$. 
As before,
$\delta$ values where chosen from the set 
$\delta \in
\left\{\frac{1}{6},\frac{1}{12},\frac{1}{18},\frac{1}{24},\frac{1}{30},\frac{1}{36},\frac{1}{48}\right\}\textrm{eV}^{-1}$
(corresponding to $N_t$'s between $ 12 $ and  $96$), 
and second order polynomials were used in all cases (the full set of
$\delta$ values was only used for the cases $\lambda=0.8/1.0$).
In Figs.~\ref{fig:susB2V12M5} we have collected the extrapolated
results for the various interaction strengths, showing the full
susceptibility (top), the connected (middle) and disconnected (bottom)
parts respectively. We observe that with increasing interaction strength
the peak of the full susceptibility at the VHS becomes more and more
pronounced.  This is due to both, a corresponding rise in the connected
part at the VHS and an additional contribution from the disconnected
part (which is clearly non-zero for the interacting system).  
The peak position as well as the upper end of the conduction band are
shifted towards smaller values of $\mu$. Note that we cannot
disentangle the squeezing of the $\pi$-bandwidth from interactions and
doping here. The combined effect certainly increases with increasing
interaction strength which is qualitatively in line with
experimental observations \cite{Ulstrup:2016ha}. 
Additionally, we observe that the thermodynamically
interesting disconnected part $\chi_\textrm{dis}$ of the susceptibility
develops a second peak close to the upper end of the band  
(corresponding to the $\Gamma$ point) which is thus a purely
interaction-driven effect.  

From Figs.~\ref{fig:susB2V12M5} (top and middle) it also appears that
the connected part $\chi_\textrm{con}(\mu)$
is slightly negative at large values of $\mu$. This is clearly
unphysical. We attribute it to a residual systematic error associated with the
$\delta \to 0$ continuum extrapolations.   
We have checked that with quadratic polynomial fits the negative offset shrinks
as additional points with smaller $\delta$ are included.

\subsection{Influence of temperature}
\begin{figure}
\begin{center}
\includegraphics[width=0.97\linewidth]{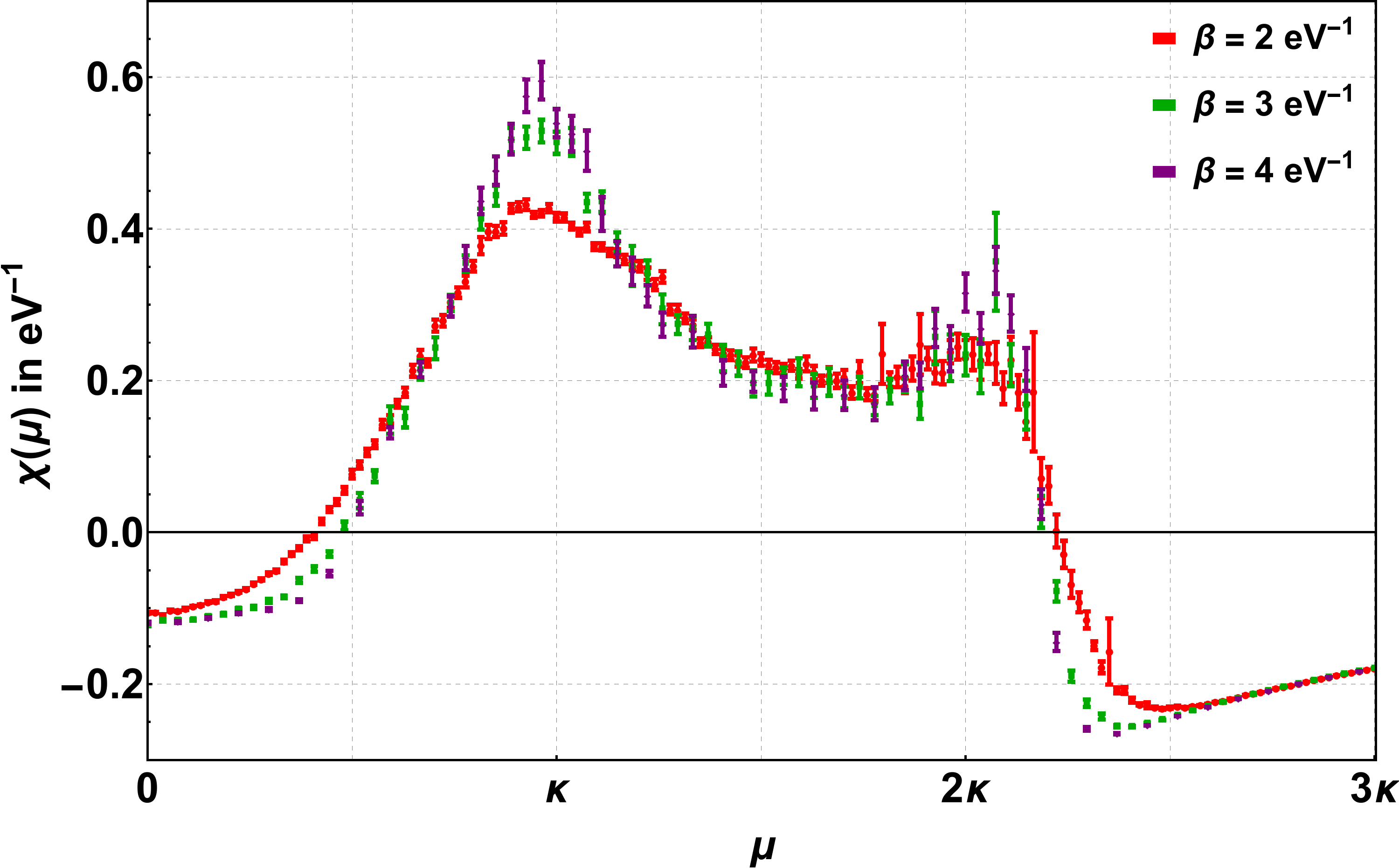}
\includegraphics[width=0.97\linewidth]{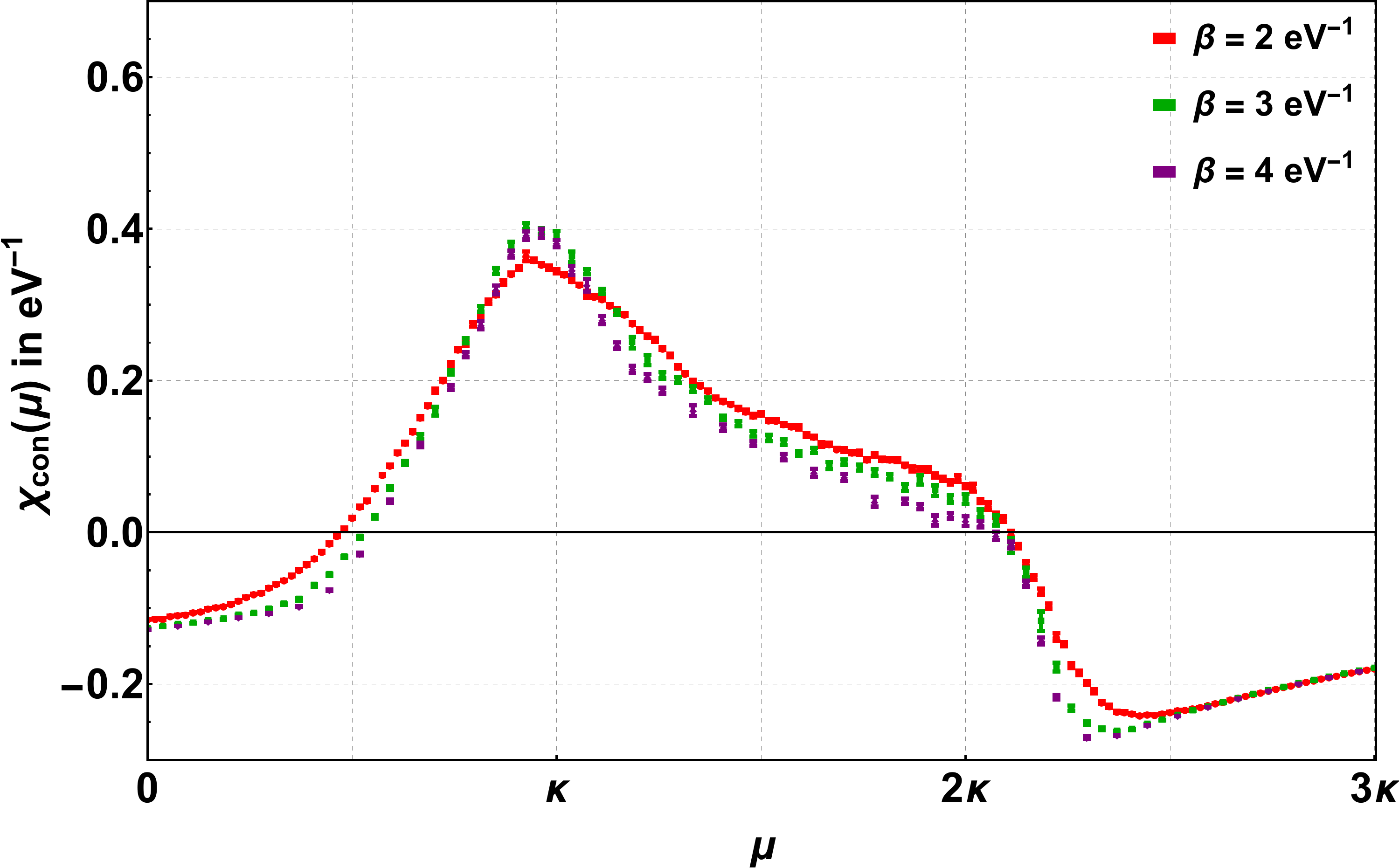}
\includegraphics[width=0.97\linewidth]{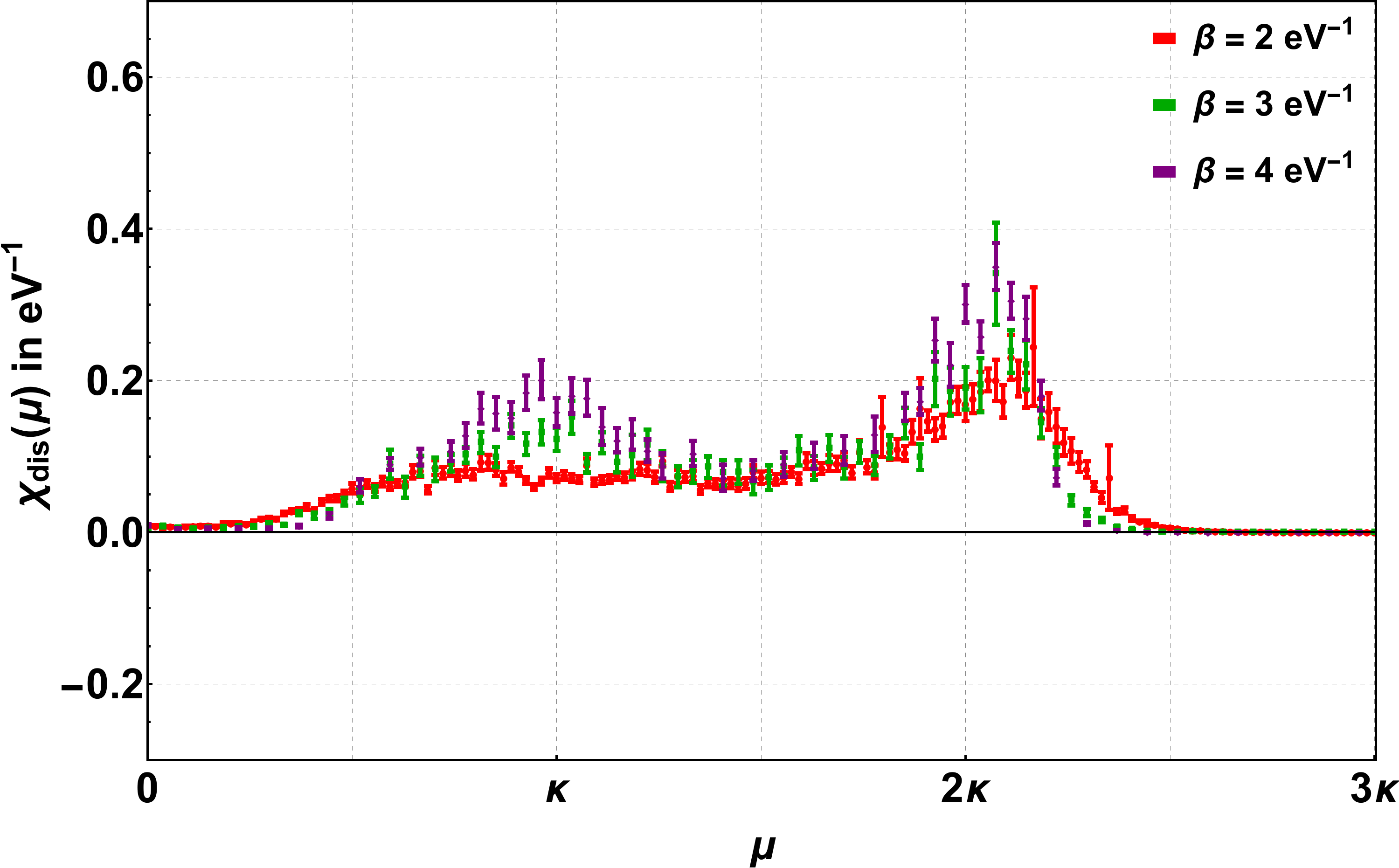}
\caption{Temperature dependence of $\chi(\mu)$ (top),
  $\chi_\textrm{con}(\mu)$ (middle) and $\chi_\textrm{dis}(\mu)$ (bottom).  
Lattice sizes scale linearly with $\beta$, such that the
displayed curves correspond to $N=12,18,24$ respectively; 
with $\delta=1/6~ \textrm{eV}^{-1}$ and $\lambda=1$ for all cases.}
\label{fig:susTEMP}
\end{center}
\end{figure}

This section focuses on the effect of electronic temperature (as no phonons
are included, the temperature of the lattice atoms is zero by definition). 
All results presented in the present section were obtained for $\lambda=1$.
Figs.~\ref{fig:susTEMP} show results 
for $\chi(\mu)$ (top), $\chi_\textrm{con}(\mu)$ (middle) and
$\chi_\textrm{dis}(\mu)$ (bottom) 
respectively over the entire range of the
conduction bands for different temperatures.
Proper lattice sizes for each temperature were chosen such that finite size
effects play no role (we first estimated the necessary lattice sizes 
from Fig.~\ref{DIFF_TEMP_OVER_VOL}, and subsequently verified 
the stability of the results under further increase of $N$ for individual
points). All results were obtained
with $\delta=1/6~ \textrm{eV}^{-1}$, which leads to a rather large
negative shift of the entire curves. Nevertheless, a clear signal can be seen
for an increase of $\chi(\mu)$, not only at the VHS, but at the upper end of
the band as well. What is even more striking is that from a comparison
of Figs.~\ref{fig:susTEMP} (middle and bottom) it is clear that
these increases are driven mainly by the disconnected parts here,
which are once more unaffected by negative offsets from the Euclidean time
discretization as observed in Sec.~\ref{sec:DISK} already. 

\begin{figure}
\begin{center}
\includegraphics[width=0.97\linewidth]{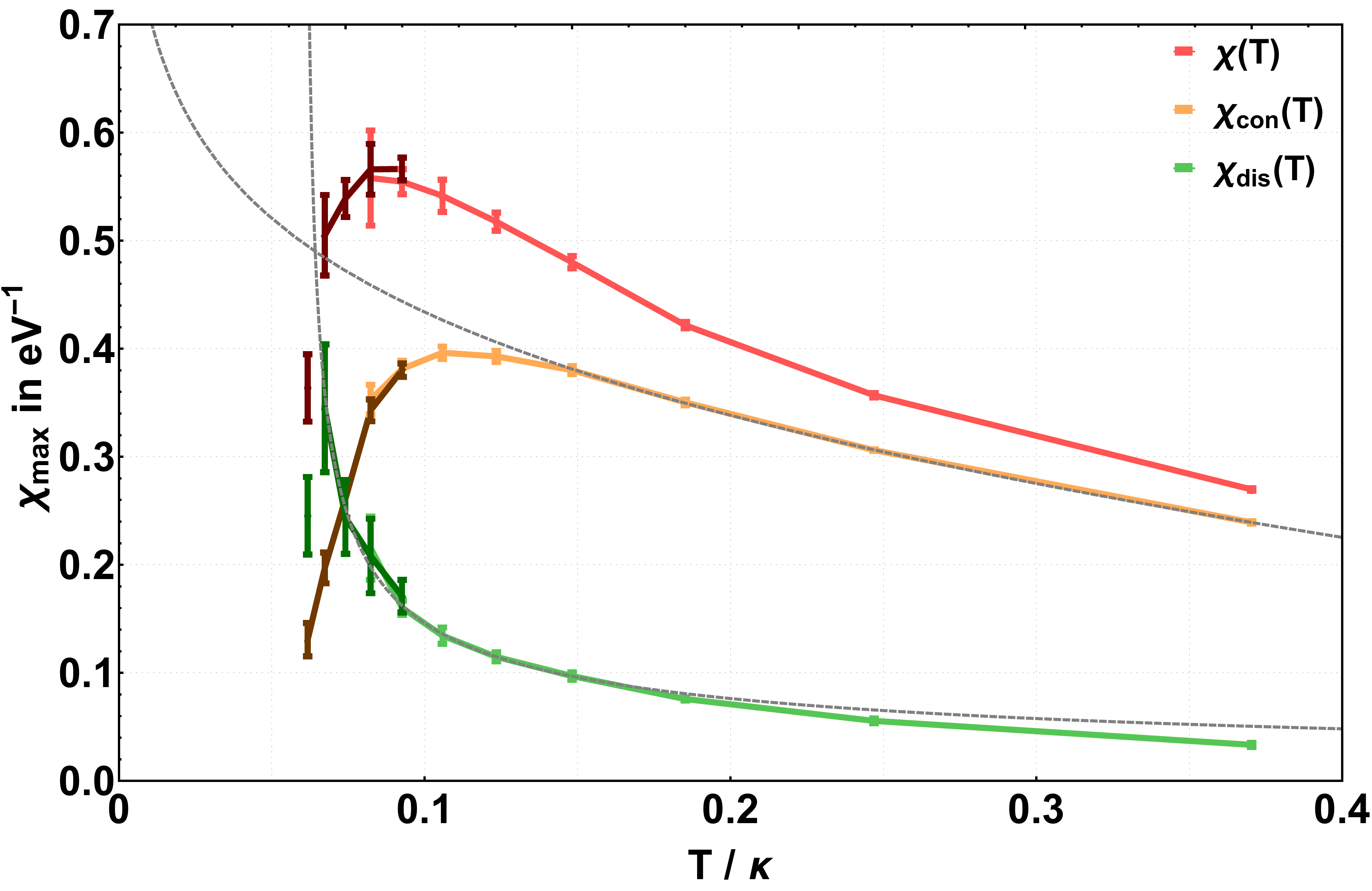}
\caption{Temperature dependence of $\chi_\textrm{max}$ in the range 
$\beta=1.0, \ldots 6.0 \textrm{ eV}^{-1}$. 
The lighter dots are from single lattices in the infinite volume limit, 
the darker dots of matching colors are obtained from average values of 
subsequent even and odd lattices. The dotted lines
are fits using Eq.~(\ref{eq:chiconfit}) for
$\chi_\textrm{con}^\textrm{max}$ and Eq.~(\ref{eq:powdivergence}) for
$\chi_\textrm{dis}^\textrm{max}$ in appropriate ranges (see text).}
\label{fig:chimaxtemp}
\end{center}
\end{figure}

To detect deviations from the temperature driven logarithmic divergence
characteristic of the neck-disrupting Lifshitz transition and described by  
Eq.~(\ref{Temppeakgl}), we simulate lattices with $\delta=1/6~
\textrm{eV}^{-1}$ in the range $\beta=1.0 \ldots 6 \textrm{
  eV}^{-1}$ in steps of $\Delta\beta=0.5 \textrm{ eV}^{-1}$. For these 
simulations we focused on the immediate vicinity of the VHS (the
position of which does not depend strongly on temperature), generating
several points in a small interval around it and using parabolic fits
to identify the peak-positions and heights of
$\chi/\chi_\textrm{con}/\chi_\textrm{dis}$. Obtaining a proper
infinite-size limit becomes increasingly problematic for lower
temperatures. In particular for $\beta = 5.0 \textrm{ eV}^{-1}$ and
larger this turned out to be too expensive to carry out in a
brute-force way. Based on the observation that the approach $N \to
\infty$ depends on the lattice parity, i.e.~whether its linear extend 
$N$ is even or odd, see Fig.~\ref{DIFF_TEMP_OVER_VOL} and the
discussion thereof,  we have thus devised a method to improve
convergence: Since even lattices overestimate 
the infinite-size limit of $\chi_\textrm{max}$ and odd lattices
underestimate it, we may expect faster convergence for average values of
two subsequent lattices of different parity. We have verified that this is
indeed so with $\beta = 4.0 \textrm{ eV}^{-1} $ and $ 4.5 \textrm{
  eV}^{-1}$, for which we compare the average values from the $N = 12$
and $13$ lattices with the converged large $N$ results in Fig.~\ref{fig:chimaxtemp}.
We then apply this averaging method for  
$\beta = 5.0 \textrm{ eV}^{-1}$, $5.5 \textrm{ eV}^{-1} $ and $6
\textrm{ eV}^{-1} $  where we have no brute-force results in the
infinite-size limit. We expect this 
method to break down close to a true thermodynamic phase transition,
as the usual finite-size scaling relations would then apply, but for
the $\beta $ values up to $ 5.5 \textrm{ eV}^{-1}$ successive average
values from $N= 11,12 $ and $11,12$ lattices still have converged
with good accuracy.\footnote{The $\beta = 6$ eV$^{-1}$ result still has somewhat 
reduced statistics compared to the others, and it is likely to be affected 
by larger systematic uncertainties from less control of finite-size effects.}

Fig.~\ref{fig:chimaxtemp} shows the temperature dependence
of the resulting infinite-size estimates for the  peak heights of
$\chi/\chi_\textrm{con}/\chi_\textrm{dis}$ obtained in this way. 
We have identified a range of $\beta =1/T$ between  $1.0$ eV$^{-1}$
and $3.0$ eV$^{-1}$ as the one over which a fit of the form 
\be
f_1(T)=a \ln\left(\frac{\kappa}{T}\right)+ b + c \, \frac{T}{\kappa}
\label{eq:logdivergence}
\ee
to the full susceptibility is possible (it breaks down if one attempts
to include lower temperatures). More interestingly, however,
the same fit to the connected part of the susceptibility alone
is consistent with $a = 3/(\pi^2\kappa ) $ for $\beta \le 2.5$ eV$^{-1}$
as predicted for the Lifshitz transition in the non-interacting system from
Eq.~(\ref{Temppeakgl}), despite the fact that we have simulated at
full interaction strength $\lambda=1$ here.
A two-parameter fit to the form
\be
 \kappa\, \chi_\textrm{con}^\textrm{max} = \frac{3}{\pi^2}
 \ln\left(\frac{\kappa}{T}\right)+b  +c \,  \frac{T}{\kappa}
 \label{eq:chiconfit}
\ee
is included in Fig.~\ref{fig:chimaxtemp}, yielding $b = 0.519(3) $ and,
for the leading $\calo(T)$ corrections in Eq.~(\ref{Temppeakgl}), $c = -
0.472(8)$ (a three-parameter fit to the form in Eq.~(\ref{eq:logdivergence})
produces $\kappa a =0.307(32)$, i.e.~a central value in $1\%$ agreement with
$\kappa a\!=\!3/\pi^2$). The result for $b$ is furthermore quite close
(within 13\%) to the constant in Eq.~(\ref{Temppeakgl}) as well, with
a discrepancy that is within the 
expected offset from the discretization $\delta = 1/6 $ eV$^{-1}$
here. We may conclude that for the larger temperatures, where the logarithmic
scaling of the peak height is observed, the behavior of the connected
susceptibility basically fully agrees with that of the non-interacting
tight-binding model in Eq.~(\ref{Temppeakgl}).  

At temperatures below $T\sim 0.15\, \kappa $ this contribution from the
electronic Lifshitz transition, which we have successfully isolated in
$\chi_\textrm{con}$, suddenly drops in the interacting theory,
however. This is contrasted by a rapid increase of the peak height of
the disconnected susceptibility  $\chi_\textrm{dis}  $ here, which
vanishes in the non-interacting limit. While $  \chi_\textrm{dis}$ is
negligible at high temperatures, it becomes the dominant contribution
to the susceptibility at $T\sim 0.07 \, \kappa$. In fact, we find that for 
$\beta \geq 2.5\textrm{ eV}^{-1}$ (corresponding to $T \le 0.15 \, \kappa$), $\chi_\textrm{dis}^\textrm{max} $
is well described by the model
\be
f_2(T)=k \left|\frac{T-T_c}{T_c} \right|^{-\gamma}~,
\label{eq:powdivergence}
\ee
resulting in the following fit parameters:
\begin{center}
\vspace{2mm}
\begin{tabular}{|c|c|c|c|}
\hline
$\beta_c~[\textrm{eV}^{-1}]$ &  $T_c~ [\kappa]$ & $\gamma~$ &$k~
    [\textrm{eV}^{-1}]$ \\ 
\hline
6.1(5) & $0.060(5) $ & $0.52(6)$ & $0.12(1)$ \\ 
\hline
\end{tabular}
\vspace{2mm}
\end{center}
The emerging peak in $\chi^\textrm{max}_\textrm{dis}(T) $
around $\beta \approx 6$ eV$^{-1}$ is thus consistent
with a powerlaw divergence indicative of a thermodynamic phase
transition at non-zero $T_c$. Despite our efforts to produce reliable
estimates for the infinite-size limits, we must expect, however, that
there are still residual finite-size effects in the points closest to
$T_c$, especially in the case of a continuous transition with a
diverging correlation length. Nevertheless, the case for a powerlaw
divergence at a finite temperature seems rather compelling here. All
attempts to model $\chi_\textrm{dis}^\textrm{max}(T)$ using a
logarithmic increase as in Eq.~(\ref{eq:logdivergence}) were certainly
unsuccessful, so that our conclusion seems qualitatively robust and
significant.   

The two most important observations are: (a) we observe good evidence of
a finite transition temperature $T_c>0 $ from the behavior of the
disconnected susceptibility as an indication of the proximity to a
thermodynamic phase transition as temperatures approach this
$T_c\approx 0.06 \, \kappa $ 
from above. (b) While the scaling exponent $\gamma \approx 0.5$ might
also be interpreted as an indication of a reshaping of the saddle points in
the single-particle band structure by the inter-electron interactions 
according to Eq.~(\ref{saddlereshape}) with an exponent $\alpha
\approx 4$ as discussed in Sec.~\ref{sec:TBLI},\footnote{As such it
  would be at  odds with the scenario of completely flat bands (the
  large-$\alpha$ limit).} because of the non-zero $T_c$
 it does not have this simple description in terms 
of independent quasi-particles with modified single-particle energies, however.  
Rather, it resembles critical behavior in the vicinity of a second-order 
phase transition. This is in line with our observation that 
here it arises in the disconnected susceptibility as mentioned above. 

\subsection{Antiferromagnetic spin-density wave susceptibility}

\begin{figure}[t]
\begin{center}
\includegraphics[width=0.97\linewidth]{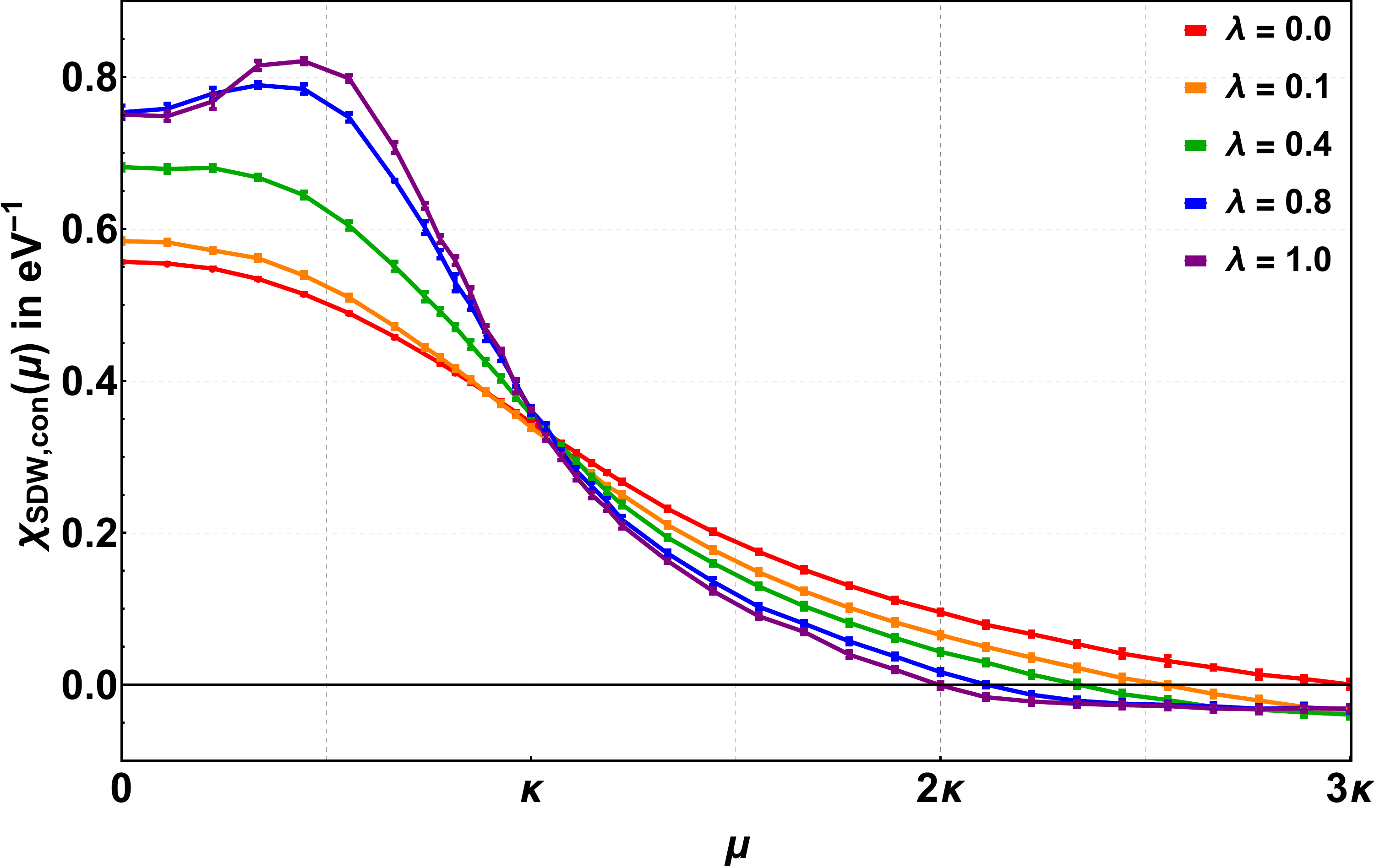}
\includegraphics[width=0.97\linewidth]{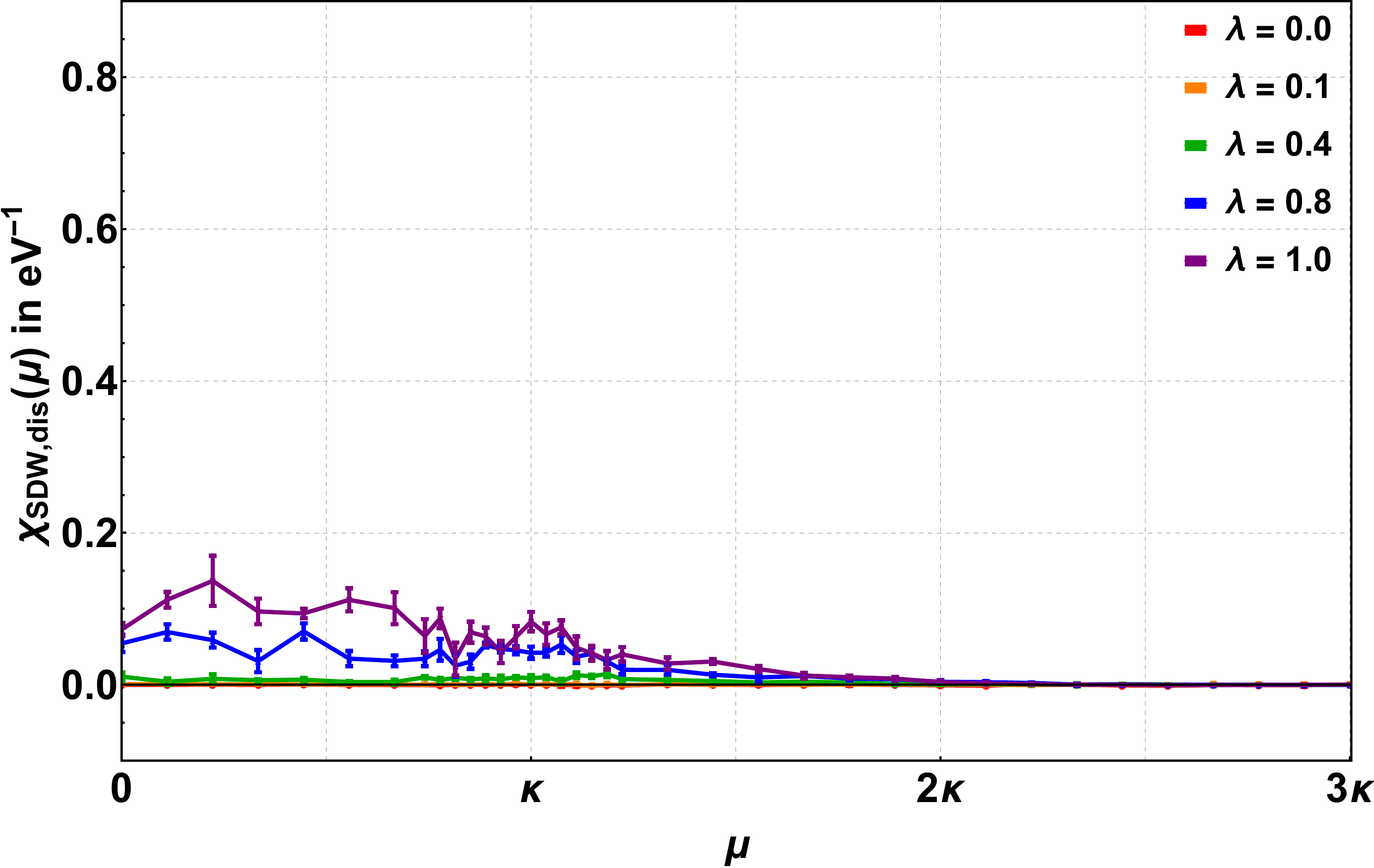}
\caption{$\chi^\textrm{sdw}_\textrm{con}(\mu)$ (top) and $\chi^\textrm{sdw}_\textrm{dis}(\mu)$ (bottom) 
for $\beta=2\textrm{ eV}^{-1}$, $N = 12$ at different interaction strengths. 
All displayed points are quadratic $\delta \rightarrow 0$ extrapolations
from simulations at non-zero $\delta$. }
\label{fig:susSDWB2V12M5}
\end{center}
\end{figure}

\begin{figure}[t]
\begin{center}
\includegraphics[width=0.97\linewidth]{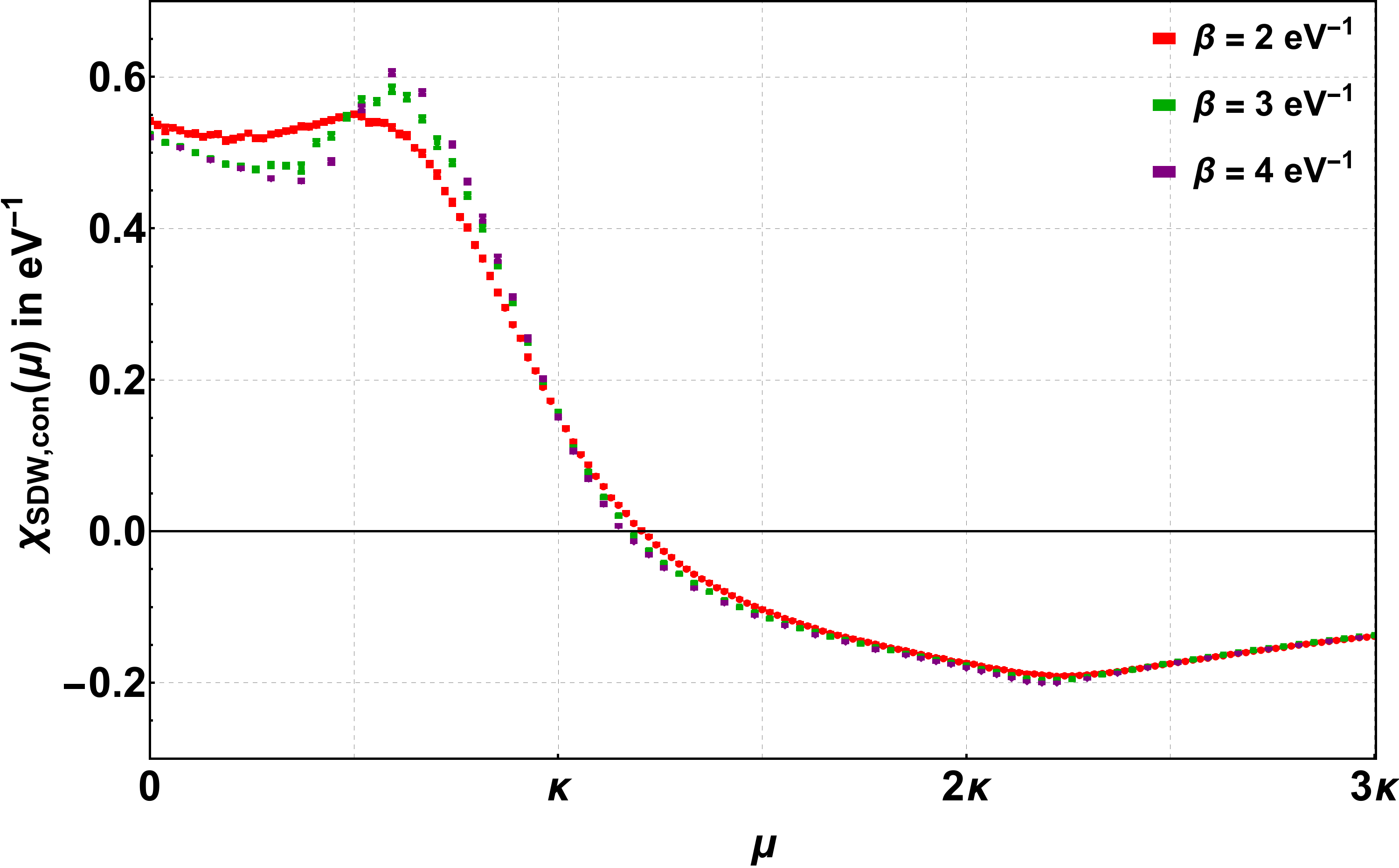}
\includegraphics[width=0.97\linewidth]{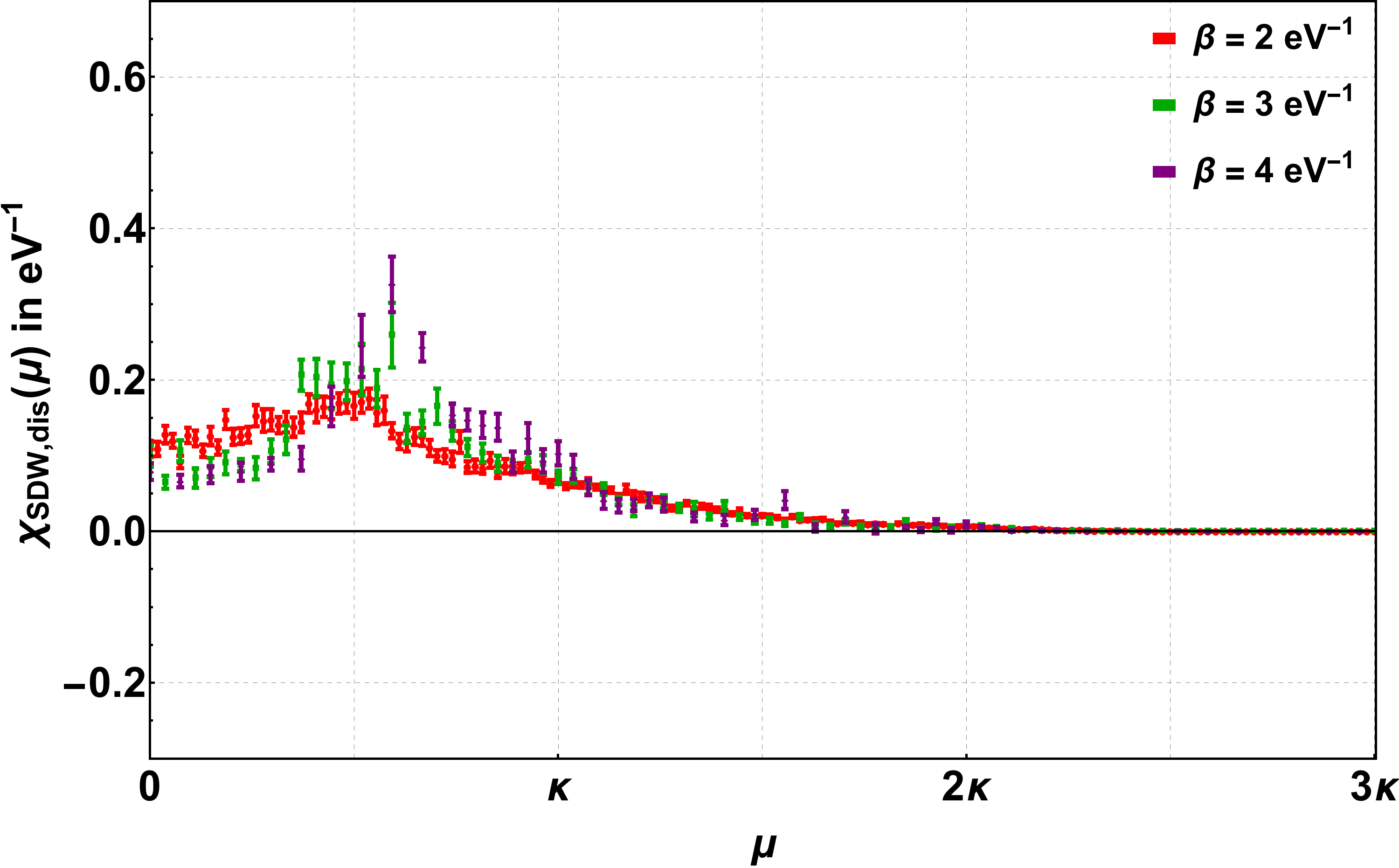}
\caption{Temperature dependence of $\chi^\textrm{sdw}_\textrm{con}(\mu)$ (top)
and $\chi^\textrm{sdw}_\textrm{dis}(\mu)$ (bottom) for $\delta=1/6~
\textrm{eV}^{-1}$ and $\lambda=1$.  
Lattice sizes scale linearly with $\beta$, such that the
displayed curves correspond to $N=12,18,24$ respectively.}
\label{fig:susSDWtemp}
\end{center}
\end{figure}

As explained in Sec.~\ref{sec:Setup} we have used here for purely computational
reasons a sublattice $s$ and spin-staggered mass $m_s = (-1)^s\, m$ in order
to regulate the low-lying eigenvalues of the fermion matrix near half
filling. This has the effect of introducing a small gap around the
Dirac points in the single-particle energy bands by triggering an
antiferromagnetic order in the ground state. For the interaction strengths
$0\le\lambda \le 1$ considered here, this order will disappear in the
limit $m\to 0$ because suspended graphene with $\lambda =1$ remains in
the semimetal phase, which has been established experimentally
\cite{Elias:2011} as well as in our present HMC simulation  
setup \cite{Ulybyshev:2013swa,Smith:2014tha}. 
 
Nevertheless, we have also measured the corresponding susceptibility
$\chi^\textrm{sdw}(\mu)$ for the antiferromagnetic spin-density
fluctuations here. While we expect no singularity at
half filling, we were particularly interested in its behavior at
finite $\mu$ in our present study. With the 
same splitting into connected and disconnected contributions,
cf.~Eqs.~(\ref{eq:sdwsusc}), our main observations are the following:
The systematics for discretization errors are completely analogous to
what was discussed above (a shift of the connected part which is
nearly independent of $\mu$ and almost no effect on the
disconnected part). As above, in Figs.~\ref{fig:susSDWB2V12M5} we
again first show the continuum extrapolated results at high temperature
$\beta = 2$ eV$^{-1}$ where this is still affordable. 
We observe an increase of $\chi^\textrm{sdw}(\mu=0)$ at half filling
with increasing interaction strength as expected. However, in addition
to this, a peak appears to form at finite $\mu$ for the larger values of
$\lambda$, mainly in $\chi^\textrm{sdw}_\textrm{con}$ but to some 
extend also visible in $\chi^\textrm{sdw}_\textrm{dis}$ which again
vanishes in the non-interacting system of course.

This peak occurs about half way between $\mu =0 $ and the VHS in the
vicinity of $\mu = \kappa$. When the temperature is lowered, however, it
it appears to move towards the VHS while getting more and more
pronounced. This is demonstrated with the ensembles at finite
discretization $\delta = 1/6 $~eV$^{-1}$ but lower temperatures and 
maximal interaction strength $\lambda = 1$ in Figs.~\ref{fig:susSDWtemp}. 
As before, there is no negative offset from the Euclidean-time
discretization in the disconnected susceptibility which shows the
increasingly sharp peak structure at the lower temperatures
particularly well.  Whether the peaks observed in the disconnected
susceptibilities of ferromagnetic and antiferromagnetic spin-density fluctuations
eventually merge and perhaps reflect the same thermodynamic phase transition
when approaching $T_c$ certainly deserves to be further studied in
the future.  

\section{Summary and Conclusions}
\label{sec:Conclusion}

We have set out to study the effects of inter-electron interactions on
the electronic Lifshitz transition in graphene. This neck-disrupting
Lifshitz transition occurs when the Fermi-level traverses the van Hove
singularity at the M-points in the bandstructure of graphene.
To elucidate the effects of interactions we have
first discussed in detail how the Lifshitz transition is reflected
in the particle-hole susceptibility of the non-interacting system,
 where it is due to a logarithmic singularity of the density of
 states. In particular we have demonstrated how this singularity
 translates into a logarithmic growth of the susceptibility maximum,
 when viewed as a function of the chemical potential, 
 with decreasing temperature  and increasing system size.

 The detailed analytical knowledge of the behavior of the particle-hole
 susceptibility in the non-interacting system, where it agrees with
 the ferromagnetic spin susceptibility, allowed us to isolate the
 same Lifshitz behavior also in presence of strong inter-electron
 interactions where it would otherwise have swamped any signs of
 thermodynamic singularities indicative of true phase transitions.   

  To search for such signs we have simulated the
  $\pi$-band electrons in monolayer with 
 partially screened Coulomb interactions, combining realistic 
 short-distance couplings with long-range Coulomb tails, using
 Hybrid-Monte-Carlo. This requires a chemical potential with
 a spin-dependent sign to circumvent the fermion-sign problem, however.
 We were therefore led to compare the ferromagnetic spin susceptibility with
 that of the non-interacting system. Despite this modification 
 our results qualitatively resemble some of the experimental results
 at finite charge-carrier density. An increase of its
 peak-height due to interactions is in-line with the existence of an
 extended van Hove singularity (EVHS) as observed in ARPES experiments
 \cite{Bostwick:2010as}. Likewise, we observe band structure renormalization
 (narrowing of the widths of the $\pi$-bands) due to interactions and
doping \cite{Ulstrup:2016ha} here as well. A possibly interesting new
feature of our results is a second peak in the spin susceptibility
$\chi (\mu)$ which arises near the upper end of the band. Whether this
is due to some form of condensation of quasi-particle pairs near the
$\Gamma$-points, 
which might happen because the Fermi levels of the different
spin-components were shifted in opposite directions, remains to be
further studied. 

The electronic Lifshitz transition itself is reflected in the connected part
of the susceptibility $\chi_\textrm{con}(\mu)$ which diverges
logarithmically in the $T \to 0$ limit when $\mu $ is at the van Hove
singularity. In the non-interacting system,   $\chi(\mu) =
\chi_\textrm{con}(\mu)$ and  $\chi_\textrm{dis}(\mu) =0 $.
With interactions, on the other hand one has  $\chi(\mu) = 
\chi_\textrm{con}(\mu) + \chi_\textrm{dis}(\mu) $.  Interestingly,
however, for higher temperatures where  $\chi_\textrm{dis}(\mu) $ is
comparatively small, the behavior of 
$\chi_\textrm{con}(\mu)$ remains precisely the same as in the
non-interacting case. The electronic Lifshitz transition is entirely
encoded in  $\chi_\textrm{con}(\mu)$. Upon its subtraction from the
full susceptibility one is left with  $\chi_\textrm{dis}(\mu) $  which
is moreover expected to be the relevant part in search for a
thermodynamic singularity reflecting a phase transition. 

In fact, our simulations provide evidence of such a thermodynamic
singularity, our results are consistent with a power-law divergence of
$\chi_\textrm{dis}$ at an electron temperature of about $T_c \approx
0.16 \textrm{ eV} \approx 0.06\, \kappa $,   
which suggests that the Lifshitz transition is replaced in the
interacting theory by a true quantum phase transition below $T_c$, and
hence for $T\to 0$ with $\mu$ as the control parameter. Without
identifying and isolating the Lifshitz behavior in
$\chi_\textrm{con}(\mu) $ it would not have been possible to observe
this with our present computational resources (we have already 
invested several hundreds of thousands of GPU hours in this project).  
The thermodynamic singularity is basically not visible in our present 
data for the full susceptibility although it will eventually dominate, 
sufficiently close to $T_c$, of course. 
 
There are a number of possible directions for future work on the
VHS. The most straightforward albeit expensive extension would be an
analysis of the critical scaling close to $T_c$. Furthermore, of direct
practical interest would be a comparison of susceptibilities
associated with different types of ordered phases such as that of the
antiferromagnetic spin-density wave order parameter studied as a first
example at the end of the last section, or superconducting phases
(e.g.~chiral superconductivity \cite{Chubukov:2012as}). It should in
principle be possible to identify the dominant instability of the VHS
and a corresponding pairing channel. 

Since the relevance of electron-phonon couplings at the VHS was demonstrated 
experimentally \cite{McChesney:2007uh}, a quantitatively exact result
should only be expected when phonons are accounted for. Furthermore,
as was demonstrated e.g.~in Ref.~\cite{PhysRevB.86.020507}, 
deviations from exact Fermi-surface nesting have a profound impact on
the competition between ordered phases. This implies that for a
realistic description the inclusion of higher order hoppings, 
which suffer from a fermion-sign problem, will be necessary. For this
reason, and due to the obvious fact that finite spin and
charge-carrier densities have different ground states, there is a solid
motivation for efforts towards dealing with the sign problem. As the
Hubbard field introduced in this work has a much simpler structure
than a non-Abelian gauge theory, it is conceivable that some of the
more recent developments
\cite{Langfeld:2014nta,Huffman:2013mla,Mukherjee:2014hsa,Alexandru:2016ejd} 
mentioned in Sec.~\ref{sec:HMCfsp} will turn
out to be useful in this context.  

\section*{Acknowledgments}
This work was supported by the Deutsche Forschungsgemeinschaft (DFG)
under grants BU 2626/2-1 and SM~70/3-1. 
Calculations have been performed on GPU clusters at the Universities of 
Giessen and Regensburg. P.B.~is also supported by a
Sofia Kowalevskaja Award from the Alexander von Humboldt foundation.

\end{document}